\documentclass[aps,prb,longbibliography,showpacs,floatfix,superscriptaddress,twocolumn]{revtex4-1}

\usepackage{graphicx,color}
\usepackage{amsfonts}
\usepackage[figuresright]{rotating}
\usepackage{amssymb}
\usepackage{amsmath}
\usepackage{mathtools}
\usepackage{psfrag}
\usepackage{subfigure}
\usepackage{multirow}
\usepackage{tabularx}
\usepackage{textcomp}
\usepackage{units}
\usepackage{lipsum}
\usepackage{soul}
\usepackage{titlesec}
\usepackage{times}
\usepackage{hyperref}
\usepackage{enumitem}
\usepackage{ulem}

\DeclareMathAlphabet\mathbfcal{OMS}{cmsy}{b}{n}

\titlespacing*{\section}
{0pt}{1ex plus .25ex}{1ex plus 1ex}
\titlespacing*{\subsection}
{0pt}{1ex plus .25ex}{1ex plus 1ex}
\titlespacing*{\subsubsection}
{0pt}{1ex plus .25ex}{1ex plus 1ex}

\titleformat{\section}
 {\fontsize{10}{15}\bfseries }
  {\thesection}
  {1em}
  {}

\titleformat{\subsection}
  {\bfseries }
  {\thesubsection}
  {2em}
  {}

\titleformat{\subsubsection}
  {\itshape}
  {\thesubsubsection}
  {2em}
  {}

\def\beq{\begin{eqnarray}}
\def\eeq{\end{eqnarray}}

\let\baraccent=\= % rename builtin command \= to \baraccent
\renewcommand{\=}[1]{\stackrel{#1}{=}} % for putting numbers above =
\newcommand{\bk}{\boldsymbol{k}} % bold k
\newcommand{\mc}[1]{\mathcal{ #1}} % mathcal
\usepackage{xcolor}
\newcommand{\bl}[1]{\textcolor{black}{#1}} % blue text
\newcommand{\org}[1]{\textcolor{black}{#1}} % orange text

\makeatletter

\titleclass{\subsubsubsection}{straight}[\subsection]

\newcounter{subsubsubsection}[subsubsection]
\renewcommand\thesubsubsubsection{\thesubsubsection.\arabic{subsubsubsection}}
\titleformat{\subsubsubsection}
   {\normalfont\normalsize\itshape \centering}{\thesubsubsubsection}{1em}{}
\titlespacing*{\subsubsubsection}
{0pt}{1ex plus .3ex}{1ex plus .3ex}
\makeatletter
\renewcommand\paragraph{\@startsection{paragraph}{5}{\z@}%
  {3.25ex \@plus1ex \@minus.2ex}%
  {-1em}%
  {\normalfont\normalsize}}
\renewcommand\subparagraph{\@startsection{subparagraph}{6}{\parindent}%
  {3.25ex \@plus1ex \@minus .2ex}%
  {-1em}%
  {\normalfont\normalsize}}
\def\toclevel@subsubsubsection{4}
\def\toclevel@paragraph{5}
\def\toclevel@paragraph{6}
\def\l@subsubsubsection{\@dottedtocline{4}{7em}{4em}}
\def\l@paragraph{\@dottedtocline{5}{10em}{5em}}
\def\l@subparagraph{\@dottedtocline{6}{14em}{6em}}
\makeatother

\begin{document}
\title{Defect bulk-boundary correspondence of topological skyrmion phases of matter}
\author{Shu-Wei Liu}
\affiliation{Max Planck Institute for Chemical Physics of Solids, Nöthnitzer Strasse 40, 01187 Dresden, Germany}
\affiliation{Max Planck Institute for the Physics of Complex Systems, Nöthnitzer Strasse 38, 01187 Dresden, Germany}

\author{Li-kun Shi}
\affiliation{Max Planck Institute for the Physics of Complex Systems, Nöthnitzer Strasse 38, 01187 Dresden, Germany}

\author{Ashley M. Cook}
\affiliation{Max Planck Institute for Chemical Physics of Solids, Nöthnitzer Strasse 40, 01187 Dresden, Germany}
\affiliation{Max Planck Institute for the Physics of Complex Systems, Nöthnitzer Strasse 38, 01187 Dresden, Germany}

\begin{abstract}
 Unpaired Majorana zero-modes are central to topological quantum computation schemes as building blocks of topological qubits, and are therefore under intense experimental and theoretical investigation. Their generalizations to parafermions and Fibonacci anyons are also of great interest, in particular for universal quantum computation schemes. In this work, we find a different generalization of Majorana zero-modes in effectively non-interacting systems, which are zero-energy bound states that exhibit a cross structure---two straight, perpendicular lines in the complex plane--- composed of the complex number entries of the zero-mode wavefunction on a lattice, rather than a single straight line formed by complex number entries of the wavefunction on a lattice as in the case of an unpaired Majorana zero-mode. These ``cross'' zero-modes are realized for topological skyrmion phases under certain open boundary conditions when their characteristic momentum-space spin textures trap topological defects. They therefore serve as a second type of bulk-boundary correspondence for the topological skyrmion phases. In the process of characterizing this defect bulk-boundary correspondence, we develop recipes for constructing physically-relevant model Hamiltonians for topological skyrmion phases, efficient methods for computing the skyrmion number, and introduce three-dimensional topological skyrmion phases into the literature.

 %\org{Bulk-boundary correspondence highlights the important feature of topological phases of matter that robust edge modes are charaterized by the topological invariant of the bulk. In this work, we study systems where pseudospin texture could form skyrmion and therefore give rise to a topological invariant as a winding number and indicate non-trivial boundary phenomena.}
 \end{abstract}
\maketitle

%\red{To do: 1. Fig. 12 must be updated and the corresponding text updated 2. Add a schematic figure of the annulus geometry and check how much more we can follow the text of Yan~\emph{et al.}~\cite{yanzhongbo2017} for calculations which generalize those. 3. Edit Discussion 4. Finalize Intro and abstract.}

%\red{smaller to do: 0.Push some figures further along in the text so that they are not all bunched together. 1. For Fig. 11 and 12, (c) and (d) should each have $|\psi(x,y)|^2$ on the vertical axis for probability density. 2. Would be good to separate Figs. 1 \& 2, 3 \& 4, 5 \& 6, and 7 \& 8. 3. Please replace BLAH for figures and equations with correct value. 3. transpose notation between supplementary and main text must be made the same.}

\section{Introduction}\label{introduction}
Discovery of the integer and fractional quantum Hall effects in the early 1980's pushed condensed matter physics beyond the Landau paradigm~\cite{chiu2016} of symmetry-breaking phases of matter, towards additional characterization of matter based on its \textit{topology}. Topological phases of matter, or those phases of matter with some observables linked to topological invariants and therefore unaffected by sufficiently small perturbations, have profound consequences. They realize quasiparticles useful for exceptionally robust quantum computation schemes~\cite{Lahtinen2017}, and their topology can manifest through bulk-boundary correspondence: a non-trivial value for a topological invariant of a material bulk yields topologically-protected, gapless boundary states~\cite{hasan2010, bhughes2013, qi2011} very promising for applications such as spintronics~\cite{He2021} and topologically-protected quantum computation~\cite{nayak2008,kitaev2001}.

Study of topological phases has greatly expanded since the discovery of the quantum spin Hall insulator, a topological phase of matter realized without an applied magnetic field or magnetic order, presenting the possibility of far broader applications than topological phases studied earlier~\cite{bernevigqsh2006}. Efforts began in earnest to search for other previously-unidentified topological phases of matter, leading to the three-dimensional topological insulator protected by time-reversal symmetry~\cite{Hasan2010rev}, and later the ten-fold way classification scheme for topological phases of matter~\cite{schnyder2008,Ryu2010}. Much more recently, efforts to expand the ten-fold way to include bosonic topological phases~\cite{Tiwari2018}, crystalline topological phases~\cite{Fu2011,Hsieh2012,Huang2017, Song2017, Cornfeld2021}, the broader set of short-range and long-range entangled topological phases~\cite{Chen2013SPT, Gong2016}, topology of various quasiparticles~\cite{Khanikaev2012, Lan2014,Susstrunk2015,Vijay2015, Dai2020, Tokuda2021} and driven~\cite{lindner2011,cayssol2013,khemani2016,Crowley2019,Wintersperger2020} and open~\cite{shen2018,gong2018,Zheng2018,kawabata2019} systems reveal the foundational role of topology in condensed matter physics.

In all of these topological phases of matter, however, topology is associated with the full set of degrees of freedom of a system. This reflects the characterization of many topological phases based on projectors onto occupied states~\cite{schnyder2008,Ryu2010}. Recent work has shown, however, that topological phases are also realized for subsets of the degrees of freedom of the occupied states. Specifically, past work showed non-trivial topology is also associated with just the spin degree of freedom (DOF) of the ground-state in symmetry-protected topological phases even when the ground-state is characterized by a larger set of degrees of freedom~\cite{Cook2022}, \textit{independent of} the previously-known topology associated with the full set of degrees of freedom of the ground-state.

These topological phases of the spin degree of freedom, known as topological skyrmion phases of matter, serve as momentum-space duals of the widely-studied skyrmion magnetic orders in real-space~\cite{roessler2006,muhlbauer2009,romming2013,fert2017}. As these skyrmions are realized in momentum-space and their symmetry-protection is robust against crystal-field splitting, however, they are topological in a stricter sense, exhibiting a bulk-boundary correspondence upon performing a partial trace over all degrees of freedom of a system except spin. As these topological phases are protected by a generalized particle-hole symmetry present in centrosymmetric superconductors, such signatures may be important for experiments and understanding of exotic mechanisms of superconductivity. The consequences of this non-trivial topology are even more significant, however, yielding novel topological phase transitions even in effectively non-interacting systems while the symmetry protecting this spin topology is preserved, as the topology is protected by a spin magnitude gap rather than an energy gap. Thus, given that the topological skyrmion phases are experimentally-relevant and generalize foundational concepts of topological condensed matter physics, better understanding of these phases will yield valuable insights.

%Additionally, while topology of strictly spin has been documented thus far, topology of other degree of freedom subsets is also anticipated.

%\org{The study of topological phases of matter has been an extensive and prominent branch of condensed matter physics. In these systems, phase transition takes the form of switching topological invariants, instead of breaking any symmetry as in the Ginzburg-Landau-Wilson formalism. Amongst many achievements in this field in the last decades, one cornerstone is the Chern insulator, a two-dimensional topological phase of matter with broken time-reversal symmetry. The topological invariant (the Chern number) is equal to the number of chiral edge states, demonstrating the bulk-boundary correspondence. The Chern insulator plays an important role in the literature because it lays the ground for other types of topological phases such as quantum spin Hall insulator, Weyl semimetal, Floquet topological phases, non-Hermitian topological phases and many more. With the first Chern insulator, the Haldane model for graphene, being established for more than thirty years, many efforts have followed suit both theoretically and experimentally to seek further generalization in order to broaden our understanding of the field.}

In this work, we examine these topological  skyrmion phases of matter, revealing they exhibit a second kind of bulk-boundary correspondence due to the momentum-space spin textures trapping defects, while also introducing the three-dimensional topological skyrmion phases of matter into the literature. These results agree with past work on topological classification of defects by Teo and Kane~\cite{teo2010}, but there are surprising additional consequences of topological defects in momentum-space. Past work shows unpaired Majorana zero-modes may be realized for certain boundary conditions when spin skyrmions are realized in momentum-space for the Hopf insulator\cite{moore2008,DengHopf2013,LiuHopf2017,yanzhongbo2017}. We extend these results to the topological skyrmion phases of matter, showing these momentum-space spin skyrmions realize more exotic topologically-protected zero-modes. In identifying this second defect-based bulk-boundary correspondence of the topological skyrmion phases of matter, we provide strong evidence that the consequences of this physics, including the type-II topological phase transition of the topological skyrmion phases, must be explored in detail to better understand the foundations of topological condensed matter physics.

The overview of the manuscript is as follows: in Sec.~\ref{2}, we review the topological skyrmion phases and then expand on earlier results by presenting methods for constructing toy models---including those corresponding to Bogoliubov de Gennes Hamiltonians for superconductors---and highly-efficient calculation of the topological invariant. We use these techniques to characterize the models numerically, studying their phase diagrams and momentum-space spin textures. In Sec.~\ref{3}, we then extend our discussion by introducing  three-dimensional topological skyrmion phases using the construction methods of Sec.~\ref{2}. Using these methods, we construct toy models for the three-dimensional skyrmion phase from the Hopf insulator~\cite{LiuHopf2017,yanzhongbo2017}. We then show these three-dimensional spin skyrmions trap momentum-space defects, which realize remarkable generalizations of unpaired Majorana zero-modes for certain boundary conditions. We characterize these so-called \textit{cross} zero-modes numerically and analytically, showing how they relate to the unpaired Majorana zero-mode. Finally, in Sec.~\ref{4}, we provide some summarizing remarks and discuss possible extensions for future research.

\section{Chiral topological skyrmion phase in two dimensions}\label{2}

\subsection{Section overview}
In this section, we first provide a brief overview of the topological skyrmion phases and the standard method for computing the topological invariant characterizing them. We then introduce the simplest, or ``atomic'', symmetry-allowed toy model Bloch Hamiltonians that can realize this non-trivial topology. Based on these atomic toy models, we then present some general design principles for constructing toy model Bloch Hamiltonians with four bands that can realize the chiral phase. We then introduce a far more efficient method for computing the topological invariant and discuss situations where it is most useful. Finally, we compute phase diagrams to characterize the topology of some canonical four-band toy models exhibiting the chiral topological skyrmion phase.

\subsection{Review of the chiral topological skyrmion phase}

The chiral topological skyrmion phase is a topological phase of matter introduced in past work~\cite{Cook2022}, which is characterized by formation of a baby skyrmion in the ground state spin expectation value texture over the two-dimensional Brillouin zone, even when systems have more degrees of freedom than just spin. The phase is realized in Bloch Hamiltonians with an even number of bands greater than two, and has previously been studied in a tight-binding model relevant to superconducting Sr\textsubscript{2}RuO\textsubscript{4}. This non-trivial topology results when the Bloch Hamiltonian is invariant under a generalized particle-hole transformation corresponding to operator $\mc{C}'$, which is the product of the particle-hole operator $\mc{C}$ (with $\mc{C}^2 = -1$), and the spatial inversion operator, $\mc{I}$. This symmetry ensures that the Hamiltonian serves as a mapping from the Brillouin zone to a space acted on by the quotient $\mathrm{Sp}(2N) / \mathrm{U}(N)$, where $\mathrm{Sp}(2N)$ is the compact symplectic group with $\mathrm{Sp}(2)$ ($N=1$) isomorphic to the special unitary group $\mathrm{SU}(2)$. One physical quantity acted upon by such a quotient is the spin angular momentum. As a result, systems with $\mc{C}'$ symmetry can realize topological phases characterized by the skyrmion number $\mc{Q}$ computed as an integral over the Brillouin zone of the normalized ground state spin expectation value $ {\boldsymbol{\hat{S}}} ({\bf k}) = {\boldsymbol{S}} ({\bf k}) / | {\boldsymbol{S}} ({\bf k}) | $, where $ {\boldsymbol{S}} ({\bf k}) = \sum_{n \in \text{filled} } \langle u_{n} ({\bf k}) | {\bf S} | u_n ({\bf k}) \rangle$ and the spin operator $\bf S$ reads
\begin{equation}
S_x = \tau_3 \sigma_1 ,
\quad
S_y = \tau_0 \sigma_2 ,
\quad
S_z = \tau_3 \sigma_3 .
\end{equation}
where here $\{\tau_i\}$ ($\{\sigma_i\}$) ($i \in \{0,1,2,3\}$ is a set of Pauli matrices acting in generalized particle-hole (spin-${1 \over 2}$) space. $\tau_0$ and $\sigma_0$ are $2 \times 2$ identity matrices.

We may express the skyrmion number as
\begin{align}
    \mc{Q} = {1 \over 4\pi}\int_{BZ}d^2k \langle \hat{\boldsymbol{S}}(\bk) \rangle \cdot \left(\partial_{k_x} \langle \hat{\boldsymbol{S}}(\bk) \rangle \times \partial_{k_y} \langle \hat{\boldsymbol{S}}(\bk) \rangle  \right).
    \label{skyrmnum1}
\end{align}

The integrand here is one component of an object we call the \textit{skyrmion curvature} $\boldsymbol{\Omega}_{\mc{Q}}$, expressed as
\begin{align}
\boldsymbol{\Omega}_{\mc{Q}} = \{\Omega_{\mc{Q},x}, \Omega_{\mc{Q},y}, \Omega_{\mc{Q},z} \},
\label{skyrmcrv}
\end{align}

where the $k$\textsuperscript{th} component is defined as

\begin{align}
    \Omega_{\mc{Q},k} = \epsilon_{ijk} \langle \hat{\boldsymbol{S}}(\bk) \rangle \cdot \left(\partial_{k_i} \langle \hat{\boldsymbol{S}}(\bk) \rangle \times \partial_{k_j} \langle \hat{\boldsymbol{S}}(\bk) \rangle  \right).
    \label{skyrmcrvpart}
    \end{align}

where $\epsilon_{ijk}$ is the Levi-Civita tensor and Einstein summation is implicit.

\subsection{Methods for constructing toy model Hamiltonians realizing the 2D chiral topological skyrmion phase}
In order to construct toy models realizing the chiral topological skyrmion phase, we require that they serve as mappings from the Brillouin zone to the space of ground state spin expectation values corresponding to non-trivial homotopy group $\pi_2\left(\mathrm{Sp}(2N) / \mathrm{U}(N) \right) = \mathbb{Z}$. As discussed in earlier work, this requires the Hamiltonian be invariant under a $\mc{C}'$ transformation~\cite{liu2017}.
For a 2$N$-band model, the $\mc{C}'$ symmetry operates on fermion operators as $\mc{C}' f_k \mc{C}'^{-1} = Jf_k^\dagger$ where
\begin{equation}
 J =
\begin{pmatrix}
0 & I_{N\times N}\\
-I_{N\times N} & 0
\end{pmatrix}.
\end{equation}

The Hamiltonian matrix representation must therefore satisfy

\begin{equation}
    J^{-1}H(\textbf{k})J = -H^*(\textbf{k}).
\end{equation}

The form of the Hamiltonian for $N=2$ (corresponding to four bands) is therefore restricted to include some or all of the ten checked terms in TABLE \ref{table:1}, expressed as Kronecker products of Pauli matrices $\tau_i \sigma_j=\tau_i \otimes \sigma_j$:

\begin{table}[h!]
\centering
\[
\begin{array}{|c|c|c|c|c|}
 \hline
 \hspace{0.3cm} & \sigma_0 & \sigma_1 & \sigma_2 & \sigma_3\\
 \hline
 \tau_0 & \times & \times & \checkmark & \times\\
 \hline
 \tau_1 & \checkmark & \checkmark & \times & \checkmark\\
 \hline
 \tau_2 & \checkmark & \checkmark & \times & \checkmark\\
 \hline
 \tau_3 & \checkmark & \checkmark & \times & \checkmark\\
 \hline
\end{array}
\]

\caption{Invariance of the Bloch Hamiltonian under the generalized particle-hole transformation $\mc{C}'$ constrains its form. The 16 possible terms, each represented by a Kronecker product of Pauli matrices  $\tau_i \sigma_j=\tau_i \otimes \sigma_j$, are each represented in a table, with the $\tau$ Pauli matrix of the term represented by a row in the table and the $\sigma$ Pauli matrix of the term represented by a column. Of the sixteen terms, ten are symmetry-allowed (corresponding to a $\checkmark$ for the corresponding entry in the table) and six are symmetry-forbidden (corresponding to $\times$ for the corresponding entry in the table).}
\label{table:1}
\end{table}

We further classify these ten terms based on how they affect the ground state spin expectation value for half-filling. To simplify the problem, we first consider each term individually, computing the spin expectation value for each $\tau_i \sigma_j$ term, of the ten terms allowed by $\mc{C}'$ symmetry, separately. We find that three of the ten terms yield non-zero values for the $x$-, $y$- or $z$-components of the spin when considered alone: specifically, $\tau_3\sigma_1$ alone yields non-zero $\langle \boldsymbol{S}_x \rangle$, $\tau_0 \sigma_2$ alone yields non-zero $\langle \boldsymbol{S}_y \rangle$, and $\tau_3 \sigma_3$ alone yields non-zero $\langle \boldsymbol{S}_z \rangle$. That is, there exists one term $\tau_{\alpha}\sigma_{\beta}$ with four eigenvectors labeled by $i
\in \{1,2,3,4 \}$ as $\{\psi_{i,\alpha \beta} \}$ satisfying the eigen problem $\tau_{\alpha}\sigma_{\beta} \psi_{i,\alpha \beta} = E_{i,\alpha \beta} \psi_{i,\alpha \beta}$ and eigenvalues $\{ E_{i,\alpha \beta}\}$ satisfying $E_{1,\alpha \beta} \leq E_{2,\alpha \beta} \leq E_{3,\alpha \beta} \leq E_{4,\alpha \beta}$, such that $\langle \boldsymbol{S}_{\gamma} \rangle  = \sum_{i \in \{1,2\}} \psi^{\dagger}_{i,\alpha \beta} \boldsymbol{S}_{\gamma} \psi^{\dagger}_{i,\alpha \beta} \neq 0$ for each $\gamma \in \{x,y,z\}$. As the chiral skyrmion phase requires realization of a skyrmionic spin texture in the Brillouin zone and therefore finite $x$-, $y$- and $z$-components of the ground state spin texture somewhere in the Brillouin zone, the simplest toy model for a non-trivial chiral skyrmion phase must therefore include these three terms. This furthermore suggests that the simplest way to construct various four-band Bloch Hamiltonians restricted to these three terms is as
 \begin{align}
     \mc{H}(\bk) &= d_1(\bk) \tau_3\sigma_1 + d_2(\bk) \tau_0 \sigma_2 + d_3(\bk) \tau_3 \sigma_3,
     \label{atomicham}
     \end{align}
where $\{d_i(\bk)\}$ are the momentum-dependent functions of a two-band model for a Chern insulator expressed as $\mc{H} = d_1(\bk) \sigma_1 + d_2(\bk) \sigma_2 + d_3(\bk) \sigma_3$.

The other seven symmetry-allowed terms may be classified depending on how they alter the $x$-, $y$-, or $z$-components of the ground state spin texture. Specifically, a spin component can be finite when one of these seven terms is multiplied by the same constant as one of the three essential terms, or zero. We can therefore construct TABLE \ref{table:2}, marking each of the additional seven terms as yielding $F$ (finite) or $Z$ (zero) spin-component when it is multiplied by the same prefactor as one of the three essential terms.
\begin{table}[h!]
\centering
\[
\begin{array}{|c|c|c|c|c|c|c|c|}
 \hline
 \hspace{0.3cm} & \tau_1 \sigma_0 & \tau_1 \sigma_1 & \tau_1 \sigma_3 & \tau_2 \sigma_0 & \tau_2 \sigma_1 & \tau_2 \sigma_3 & \tau_3 \sigma_0\\
 \hline
 \tau_3\sigma_1 & F & F & Z & F& F & Z & Z\\
 \hline
 \tau_0 \sigma_2 & Z & F & F & Z& F & F & Z\\
 \hline
 \tau_3 \sigma_3 & F & Z & F & F& Z & F & Z\\
 \hline
\end{array}
\]

\caption{In Eqn.~\eqref{atomicham}, each component of the $d$-vector $\boldsymbol{d}(
\boldsymbol{k}) = \langle d_1(\boldsymbol{k}), d_2(\boldsymbol{k}), d_3(
\boldsymbol{k}) \rangle$ multiplies a single Kronecker product of Pauli matrices (
$\tau_3 \sigma_1$, $\tau_0\sigma_2$, or $\tau_3 \sigma_3$, respectively). We may instead multiply each component of the $d$-vector by multiple symmetry-allowed Kronecker product terms. At the simplest level, we multiply instead by a sum of one of  $\tau_3 \sigma_1$, $\tau_0\sigma_2$, or $\tau_3 \sigma_3$ (which correspond to each row of the table) and one of the remaining seven terms (which correspond to each column of the table). When the row term and column term are each multiplied by the same pre-factor, they yield either zero for one component of the ground state spin vector, such that the magnitude of the ground state spin is zero somewhere in the Brillouin zone (corresponding to $Z$ in the table), or finite spin-component (corresponding to $F$ in the table). The latter scenario ensures there is a choice of $d$-vector yielding finite ground-state spin magnitude for each momentum $\boldsymbol{k}$ in the Brillouin zone, such that the skyrmion number $
\mc{Q}$ is well-defined.}
\label{table:2}
\end{table}
This forms the basis for constructing toy model Hamiltonians realizing the 2D chiral topological skyrmion phase, but does not exhaust all possibilities for symmetry-allowed toy model Bloch Hamiltonians with four bands.

\subsection{Relation between the Chern number and the Skyrmion number in a special model}

Here we investigate the relation between the Chern number and the skyrmion number in a special case.

\bigskip

The Hall conductivity in an ideal lattice evaluated directly from the Kubo formula reads
\begin{align}
\sigma_{x y} (\omega) = \frac{e^2 \hbar}{V}
& \sum_{{\bf k}, m \neq n}
( f_{m {\bf k}} -f_{n {\bf k}} ) \times
\nonumber\\
& \frac{{\rm Im}\, \big[ \langle u_{m} ({\bf k}) | v_x | u_{n} ({\bf k}) \rangle \langle u_{n} ({\bf k}) | v_y | u_{m} ({\bf k}) \rangle \big] }{ ( \epsilon_{n {\bf k}} - \epsilon_{m {\bf k}} ) (  \epsilon_{n {\bf k}} - \epsilon_{m {\bf k}} - \hbar \omega - i \eta )  } ,
\end{align}
where $m, n$ are band indices, $\omega$ and $\eta$ are set to zero in the dc clean limit, the velocity operator $v_a = \partial H ({\bf k}) / \hbar \partial k_a  $ $(a = x,y)$ in Bloch basis, and $V$ is the volume of the crystal in the real space.

For a $4$-band model described by $H ({\bf k}) = \sum_\alpha  d_\alpha ({\bf k}) \Gamma^\alpha$, we can express the velocity operator explicitly as
$v_a  =  \sum_\alpha [ \partial_{k_a} d_\alpha  ({\bf k}) ] \Gamma^\alpha / \hbar $.
The hermicity of $H ({\bf k})$ mandates that $ \Gamma^\alpha \in \{ \tau_i \otimes \sigma_j ~ | ~ 0 \leq i,  j \leq 3 \} $.
And it follows that the matrix element for the velocity operator reads
\begin{equation}
\langle u_{n} ({\bf k}) |  v_a | u_{m} ({\bf k}) \rangle =
\sum_\alpha [ \partial_a d_\alpha ({\bf k}) ]  \langle \Gamma^\alpha ({\bf k}) \rangle_{nm} / \hbar ,
\quad
(a = x,y)
\end{equation}
where $\langle \Gamma^\alpha ({\bf k}) \rangle_{nm} \equiv \langle u_{n} ({\bf k}) | \Gamma^\alpha | u_{m} ({\bf k}) \rangle$.
Using the explicit form of velocity matrix elements, the dc Hall conductivity reads
\begin{align}
\sigma_{x y}
= \frac{e^2}{h} & \Big[  \frac{1}{2\pi} \int_{\rm BZ} {\rm d} {\bf k} \sum_{ m \neq n}
\sum_{\alpha \beta}
( f_{m {\bf k}} -f_{n {\bf k}} ) \times
\nonumber\\
& [ \partial_x d_\alpha ({\bf k}) ] \, [ \partial_y d_\beta ({\bf k}) ]
\frac{ {\rm Im}\, \big[ \langle \Gamma^\alpha ({\bf k}) \rangle_{m n}
\langle \Gamma^\beta ({\bf k}) \rangle_{nm}  \big] }{ ( \epsilon_{n {\bf k}} - \epsilon_{m {\bf k}} )^2  }  \Big]
\nonumber\\
= \frac{e^2}{h} & {\cal C} ,
\end{align}
with $\cal C$ the well-known Chern number. We note that the second equation is valid only when the chemical potential is inside the gap.

\bigskip
Following Eqn.~\eqref{skyrmnum1} as the expression for the skyrmion number and focusing on the Hamiltonian Eqn.~\eqref{atomicham}, we have the following properties:
\\
(1)
\begin{align}
\langle u_{n} ({\bf k}) | {\bf S} | u_{n} ({\bf k}) \rangle = - {\bf d} ({\bf k}) / d ({\bf k}) , \hspace{2mm}n \in \{1,2\}
\nonumber\\
\langle u_{n} ({\bf k}) | {\bf S} | u_{n} ({\bf k}) \rangle = + {\bf d} ({\bf k}) / d ({\bf k}), \hspace{2mm}n \in \{3,4\}
\end{align}
where ${\bf d} ({\bf k}) = \big( d_1 ({\bf k}) , d_2 ({\bf k}) , d_3 ({\bf k}) \big)$, $d ({\bf k}) = | {\bf d} ({\bf k}) |$, and $n = 1,2,3,4$ refers to the band index;
\\
(2)
\begin{equation}
\epsilon_{1 {\bf k}} = \epsilon_{2 {\bf k}} = - d ({\bf k}) ,
\quad
\epsilon_{3 {\bf k}} = \epsilon_{4 {\bf k}} = + d ({\bf k}) ;
\end{equation}
(3)
\begin{equation}
\sum_{n = 3,4} {\rm Im}\, \big[ \langle S_a ({\bf k}) \rangle_{m n}
\langle S_b ({\bf k}) \rangle_{n m}  \big]
= - \epsilon_{a b c}
\frac{d_c ({\bf k})}{ d ({\bf k}) } ,
\quad
(m=1,2)
\end{equation}
and
\begin{equation}
\sum_{n = 1,2} {\rm Im}\, \big[ \langle S_a ({\bf k}) \rangle_{m n}
\langle S_b ({\bf k}) \rangle_{n m}  \big]
= + \epsilon_{a b c}
\frac{d_c ({\bf k})}{ d ({\bf k}) } .
\quad
(m=3,4)
\end{equation}

With the above three properties (1), (2), and (3), at half filling (bands $1,2$ are filled) we can rewrite the Chern number as:
\begin{align}
{\cal C} & = \frac{1}{2\pi} \int_{\rm BZ} {\rm d} {\bf k} \,
4 \sum_{a b}
[ \partial_x d_a ({\bf k}) ] \, [ \partial_y d_b ({\bf k}) ]
\frac{  - \epsilon_{a b c}  d_c ({\bf k}) }{ 4 \, d^3 ({\bf k})  }
\nonumber\\
& = - 2 \, \Big[ \frac{1}{4\pi} \int_{\rm BZ} {\rm d} {\bf k} \,  {\boldsymbol{\hat{S}}} ({\bf k}) \cdot \Big( \partial_{k_x} {\boldsymbol{\hat{S}}} ({\bf k}) \times \partial_{k_y} {\boldsymbol{\hat{S}}} ({\bf k}) \Big) \Big]
\nonumber\\
& = - 2 {\cal Q} .
\label{CQ relation}
\end{align}

\subsection{Efficient method for computing two-dimensional skyrmion number using skyrmion connection}

While the skyrmion number may be computed using the expression given in Eq.~\eqref{skyrmnum1}, we may also compute it using a far more efficient method based on projectors, developed for computing the Chern number in past work~\cite{fukui2005, chen2013}. The idea of this earlier work was that computing a Berry phase directly from summation of Berry curvature is more costly than computing it as a sum of infinitesimal Berry phases, each computed using a discretized form of the Berry connection. These past works found that the latter approach converged to the correct, integer-quantized values for Chern numbers using a far coarser mesh of $\boldsymbol{k}$-points in the Brillouin zone than did the former. In analogy to this approach, we want to compute the skyrmion number with a connection, rather than the skyrmion curvature defined previously. We therefore first present an efficient method for computing this connection starting from the ground-state spin expectation value texture over the Brillouin zone.

To compute this skyrmion connection, we first construct an eigenstate of effective two-band model from the numerically computed ground state spin expectation value $\langle \hat{\boldsymbol{S}}(\bk) \rangle$ of our four-band model. Writing $\langle \hat{\boldsymbol{S}}(\bk) \rangle = \{\langle \hat{\boldsymbol{S}}_x(\bk) \rangle, \langle \hat{\boldsymbol{S}}_y(\bk) \rangle, \langle \hat{\boldsymbol{S}}_z(\bk) \rangle \} = \{d_{S,x}(\bk), d_{S,y}(\bk), d_{S,z}(\bk) \}$, we may write the effective two-band model $\mc{H}_{\mathrm{eff}}$ as
\begin{align}
    \mc{H}_{\mathrm{eff}} & = d_{S,x}(\bk) \sigma_x + d_{S,y}(\bk)\sigma_y + d_{S,z}(\bk)\sigma_z.
    \end{align}
Assuming half-filling of this effective two-band model, we may then write the form of the occupied eigenstate $|\psi_{-}(\bk) \rangle$ analytically in terms of $\{d_{S,i}\}$ as

\begin{align}
|\psi_{-}(\bk) \rangle &=  \begin{pmatrix}
\sin\frac{\theta}{2}e^{-i\phi}\\
-\cos\frac{\theta}{2}
\end{pmatrix}
\end{align}
where $\theta = \arccos\frac{d_{S,z}(\bk)}{d(\bk)}$ and $\phi = \arctan\frac{d_{S,x}(\bk)}{d_{S,y}(\bk)}$.

$|\psi_{-}(\bk) \rangle$ may then be used to compute the skyrmion number as the Chern number for the lower band $\mc{C}_-$ of $\mc{H}_{\mathrm{eff}}$, given that the skyrmion number and Chern number are equal in \textit{two-band} models at half-filling.

We evaluate $\mc{C}_-$ using a discrete lattice corresponding to the discrete values of the crystal momentum in a finite sample with periodic boundary conditions. The Chern number of the lower band may then be evaluated as
\begin{equation}
\mc{C}_- = {1 \over 2 \pi} \sum_{\bk} \mathrm{Im} \ln \left( A^n_{\bk, \hat{x}} A^n_{\bk+\hat{x}, \hat{y}} A^n_{\bk+ \hat{x} + \hat{y}} A^n_{\bk+ \hat{y} , -\hat{y}} \right),
\label{skyrmnum2}
\end{equation}
where $ A^n_{\bk, \mu} = \langle \psi_- (\bk) | \psi_-(\bk + \mu \rangle $ is the Berry connection field of the two-band model and the $\{\mu\}$ are the nearest-neighbor vectors of the square momentum-space lattice. In the context of this effective two-band model, we call this the $i$\textsuperscript{th} component of the \textit{skyrmion connection} $\boldsymbol{A}_{\mc{Q}}$, given that this effective two-band model is constructed from the ground state spin texture of a model for a chiral topological skyrmion insulator phase with more than two bands.

This connection is more generally expressed, for the $-$ band, as

\begin{align}
\boldsymbol{A}_{\mc{Q}} = \{A_{\mc{Q},x}, A_{\mc{Q},y}, A_{\mc{Q},z} \},
\label{skyrmconnect}
\end{align}
with the $i$\textsuperscript{th} component expressed as

\begin{align}
    A_{\mc{Q},i} = - i \langle \psi_- | {\partial \over \partial k_i} | \psi_- \rangle .
    \label{skyrmconnpart}
\end{align}

The corresponding \textit{skyrmion curvature}, which is the curl of this skyrmion connection, is the integrand of Eq.~\ref{skyrmcrv}.

While this approach of relating the skyrmion number in a $2N$-band model to the Chern number of a two-band model to make use of additional computation methods is generally not needed at the four-band level for skyrmion textures with topological charge sufficiently small in magnitude, it becomes invaluable for higher-magnitude topological charges or skyrmion numbers in models with more bands. As we show later, this effective connection may be used to compute the three-dimensional skyrmion number as well.

\subsection{Characterization of canonical toy models}

Here, we construct some example toy models for $N=2$ chiral topological skyrmion phases in two dimensions. We first present some example Hamiltonians to illustrate the decoupling of the total Chern number and skyrmion number at the four-band level by computing phase diagrams. We then discuss specific signatures of non-trivial skyrmion number at the four-band level beyond the momentum-space skyrmionic spin texture itself: a contribution to the Hall conductivity that may be expressed in terms of a quantized skyrmion number in $\mc{C}'$-invariant models, and finite spin current signatures required for a system to possess non-trivial skyrmion number.

We first consider a model with quantized total Chern number and quantized skyrmion number, corresponding to co-existing stable Chern insulator and chiral topological skyrmion insulator phases, by restricting our Hamiltonian to only the three special terms in TABLE~\ref{table:2}. We then show that a model constructed using three of the seven other terms in the table may realize a Chern insulator phase characterized by a stable non-trivial total Chern number at half-filling but be topologically unstable according to the skyrmion number. After this, we consider toy models for superconductors written in Bogoliubov de Gennes (BdG) formalism, discussing constraints on the Hamiltonian required for it to describe topological skyrmion phases of matter in this physical setting and consider a representative case with spin singlet pairing and another case with spin-triplet pairing.

\subsubsection{Toy model with coexisting Chern insulator and skyrmion insulator phases}
\begin{figure}[t]
\centering
\includegraphics[width=0.5\textwidth]{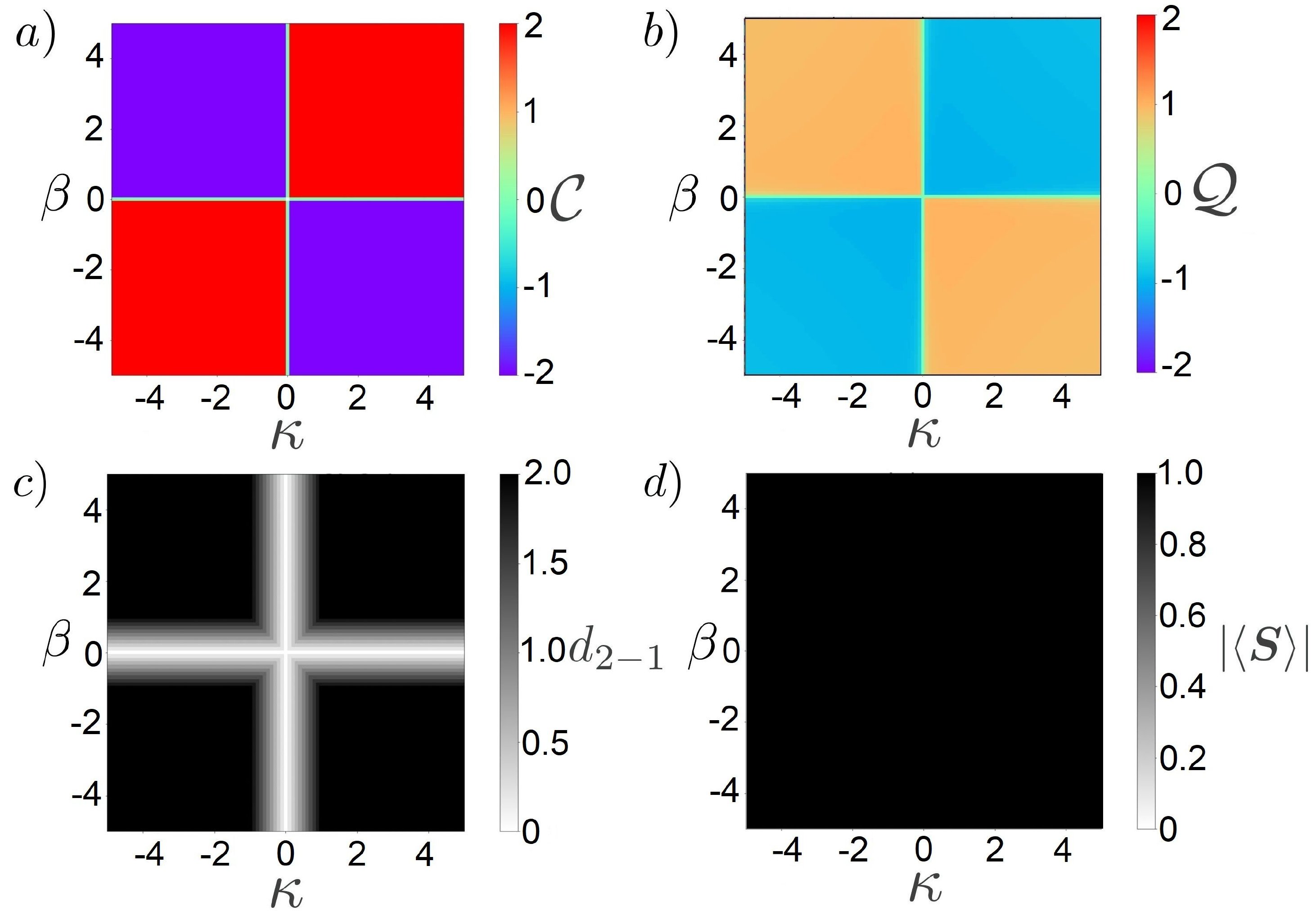}
\caption{Phase diagrams for Hamiltonian Eq.~\eqref{eqn: H1} at half-filling as a function of free parameters $\kappa$ and $\beta$ for the (a) total Chern number $\mc{C}$, (b) skyrmion number  $\mc{Q}$, (c) minimum direct energy gap between the second-lowest and third-lowest bands, $d_{2-1}$, and (d) minimum magnitude of the ground-state spin expectation value $|\langle S \rangle |$.}
\label{fig:4band1}
\end{figure}

As the three special terms mentioned above can each support a non-zero spin component, we first consider the following Bloch Hamiltonian:
\begin{align}
    \mc{H}(\bk) &= [1-\cos(k_xa)-\cos(k_ya)]\tau_3\sigma_3 +\beta \sin(k_xa) \tau_0\sigma_2 \nonumber\\
    &+\kappa \sin(k_ya) \tau_3\sigma_1.
    \label{eqn: H1}
\end{align}

\begin{figure}[t]
\centering
\includegraphics[width=0.5\textwidth]{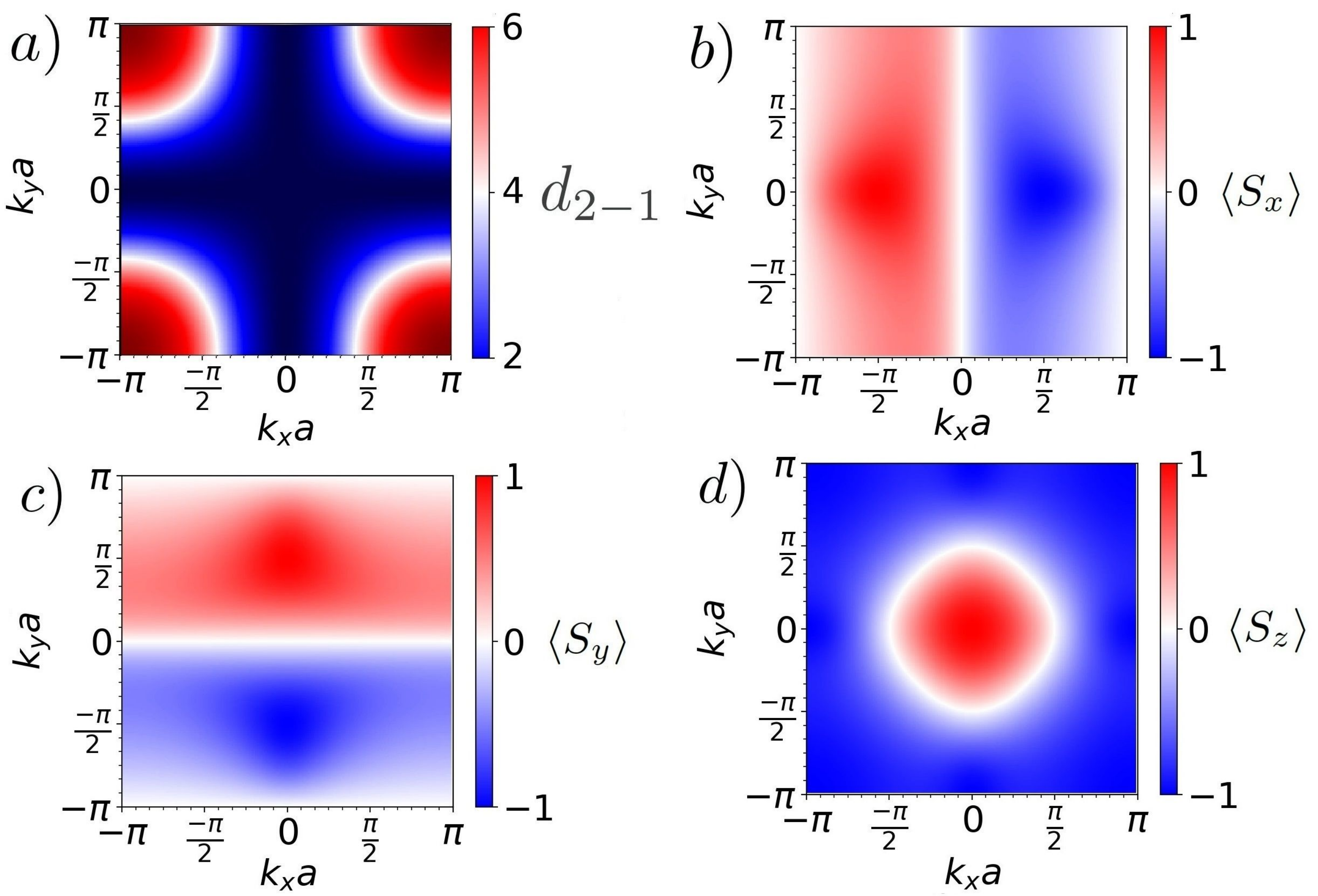}
\caption{Characterization of a representative point in the phase diagrams for Bloch Hamiltonian Eq.~\eqref{eqn: H1} shown in Fig.~\ref{fig:4band1} ($\beta=\kappa=1$) with non-trivial skyrmion number $\mc{Q}$. Minimum direct gap between the bands second-lowest and third-lowest in energy over the first Brillouin zone is shown in (a). $x$, $y$, and $z$ components of the normalized ground-state spin expectation value, $\langle S_x \rangle$, $\langle S_y \rangle$, and $\langle S_z \rangle$, over the first Brillouin zone are shown in (b), (c), and (d), respectively, and correspond to winding of the spin in agreement with the non-trivial skyrmion number value shown in Fig.~\ref{fig:4band1} (b).}
\label{fig:Sconfig1}
\end{figure}
To characterize the topology of this model, we compute the total Chern number $\mc{C}$ and skyrmion number $\mc{Q}$ assuming half filling of the bands, as well as the minimum direct gap between the bands second-lowest and third-lowest in energy over the Brillouin zone, $d_{2-1}$, and the minimum magnitude of the ground state spin expectation value over the Brillouin zone $|\langle \boldsymbol{S} \rangle|_{min}$, to obtain the phase diagrams shown in Fig.~\ref{fig:4band1}. The total Chern number $\mc{C}$, shown in Fig.~\ref{fig:4band1}(a), and the skyrmion number $\mc{Q}$, shown in Fig.~\ref{fig:4band1}(b), are each integer-quantized. We find a relationship between them of $\mc{C} = -2 \mc{Q}$, in agreement with Eqn.~\eqref{CQ relation} for the atomic model. The minimum direct gap between the second-lowest and third-lowest bands in energy, $d_{2-1}$, is shown in Fig.~\ref{fig:4band1}(c), and indicates that all topological phase transitions are type-I: they occur with a closing of a direct gap. Type-II topological phase transitions are not expected at the four-band level, due to restrictions on DOFs of the Bloch Hamiltonian other than the spin-$1/2$ DOF. We also find the minimum magnitude of the ground state spin expectation value over the Brillouin zone is finite everywhere in the phase diagram, as shown in Fig.~\ref{fig:4band1}(d), as required for well-defined and integer-quantized $\mc{Q}$.

We examine a point in the phase diagram corresponding to $\beta=\kappa=1$ in greater detail, computing the minimum direct gap between the second-lowest band in energy and the third-lowest band in energy, $d_{2-1}$, over the Brillouin zone, as well as the momentum-space spin texture, as shown in Fig.~\ref{fig:Sconfig1}. $d_{2-1}$ over the Brillouin zone is shown in Fig.~\ref{fig:Sconfig1} (a) and confirmed to be finite everywhere the Brillouin zone. It also reflects the four-fold rotational symmetry of the model. Each of the three components of the spin expectation value are shown in Fig.~\ref{fig:Sconfig1}(b), (c) and (d) respectively. They collectively show the winding of the spin vector in momentum space corresponding to a baby skyrmion.

\subsubsection{Toy model with well-defined total Chern number and ill-defined skyrmion number}

\begin{figure}[t]
\centering
\includegraphics[width=0.5\textwidth]{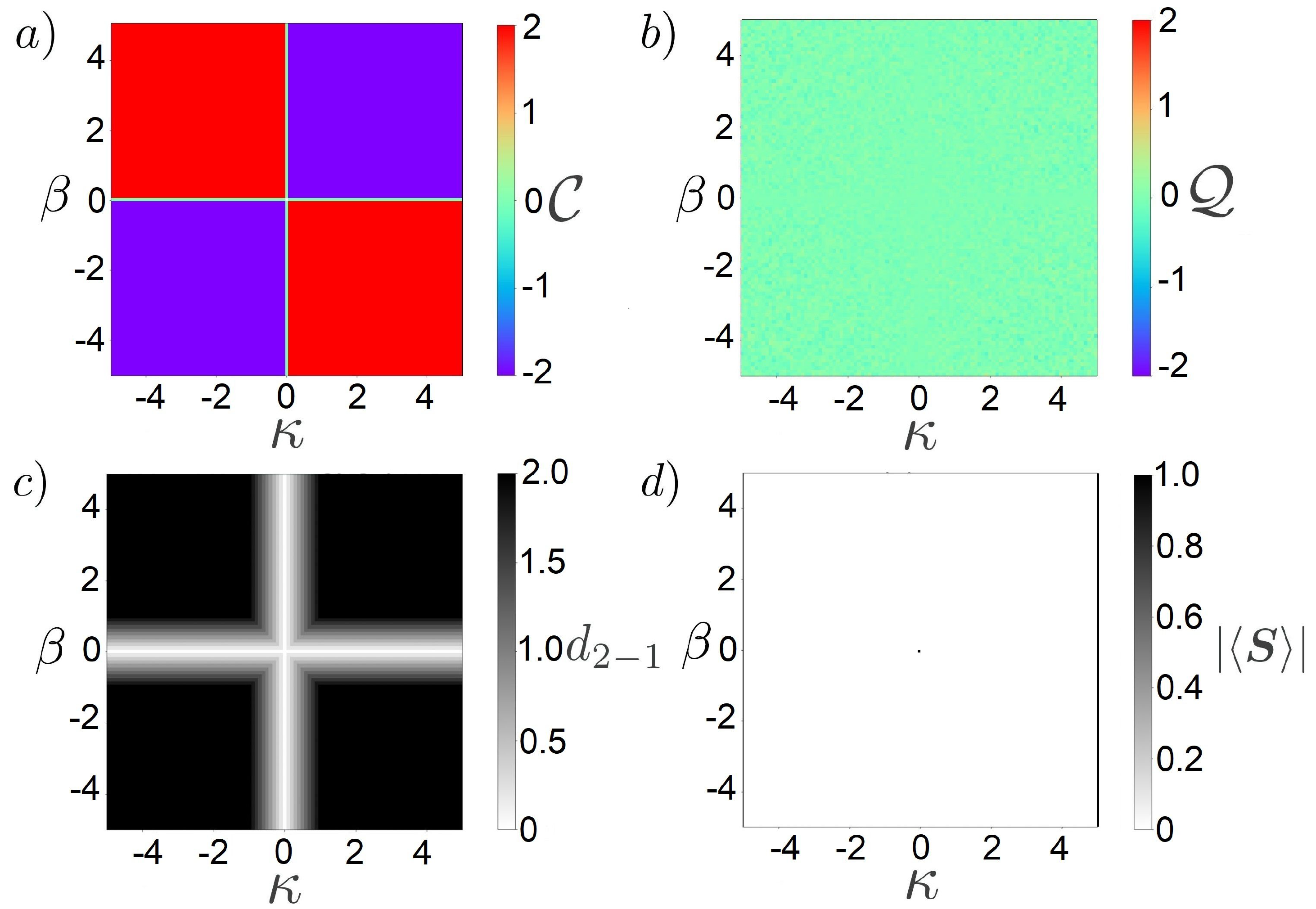}
\caption{ Phase diagrams for Hamiltonian Eq.~\eqref{eqn: H2} at half-filling as a function of free parameters $\kappa$ and $\beta$ for the (a) total Chern number $\mc{C}$, (b) skyrmion number  $\mc{Q}$, (c) minimum direct energy gap between the second-lowest and third-lowest bands, $d_{2-1}$, and (d) minimum magnitude of the ground-state spin expectation value $|\langle S \rangle |$.}
\label{fig:4band2}
\end{figure}

To further demonstrate that the total Chern number and skyrmion number are decoupled at the four-band level, we consider a model constructed from a two-band Chern insulator Hamiltonian, which has zero magnitude for the ground state spin expectation value everywhere in the Brillouin zone. We choose the following Bloch Hamiltonian:
\begin{align}
    \mc{H}(\bk) &= \left[1-\cos(k_xa)-\cos(k_ya)\right]\tau_3\sigma_3 +\beta \sin(k_xa) \tau_1\sigma_3 \nonumber\\
    &+\kappa \sin(k_ya) \tau_2\sigma_0,
    \label{eqn: H2}
\end{align}
and characterize it at half-filling by computing phase diagrams for the total Chern number $\mc{C}$, skyrmion number $\mc{Q}$, minimum direct gap between the bands second and third-lowest in energy, $d_{2-1}$, and the minimum ground state spin magnitude over the Brillouin zone, $|
\langle S \rangle |$ shown in Fig.~\ref{fig:4band2}.

\begin{figure}[t]
\centering
\includegraphics[width=0.5\textwidth]{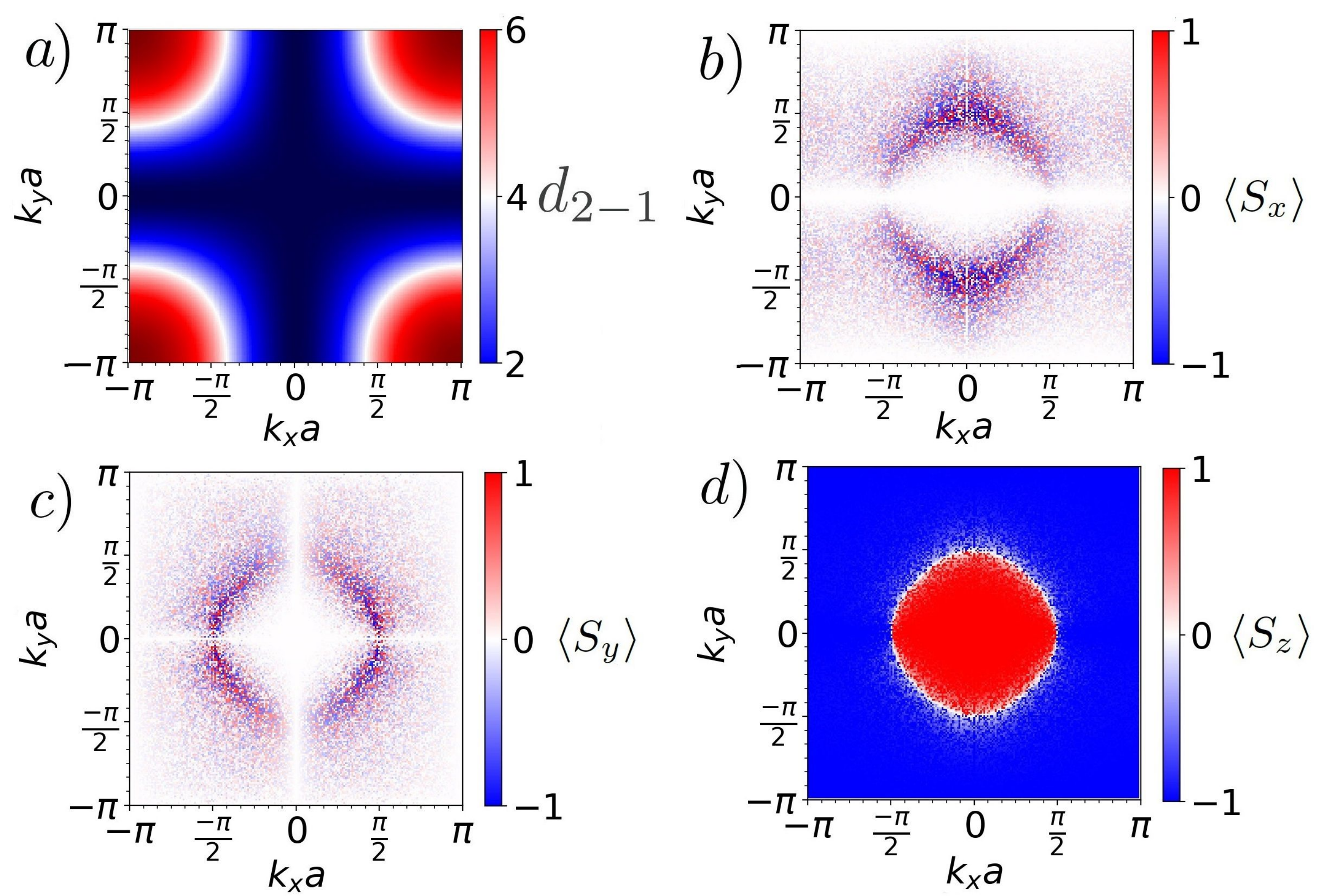}
\caption{Characterization of a representative point in the phase diagrams for Bloch Hamiltonian Eq.~\eqref{eqn: H2} shown in Fig.~\ref{fig:4band2} ($\beta=\kappa=1$) with ill-defined skyrmion number $\mc{Q}$. Minimum direct gap between the bands second-lowest and third-lowest in energy over the first Brillouin zone is shown in (a). $x$, $y$, and $z$ components of the normalized ground-state spin expectation value, $\langle S_x \rangle$, $\langle S_y \rangle$, and $\langle S_z \rangle$, over the first Brillouin zone are shown in (b), (c), and (d), respectively, and correspond to unstable skyrmion number shown in Fig.~\ref{fig:4band2} (b) due to the minimum ground state spin magnitude being zero as shown in Fig.~\ref{fig:4band2} (c).}
\label{fig:Sconfig2}
\end{figure}

The phase diagram for the total Chern number of this Hamiltonian (Eq.~\eqref{eqn: H2}) in Fig.~\ref{fig:4band2} (a) is identical to that of the total Chern number for the Hamiltonian Eq.~\eqref{eqn: H1}. However, the skyrmion number is uniformly zero as shown in Fig.~\ref{fig:4band2}(b), because the $x$ and $y$ spin components do not wind, as shown in Fig.~\ref{fig:4band2} (b) and (c), respectively. Additionally, the skyrmion number is actually ill-defined, as the minimum magnitude of the ground state spin expectation value over the Brillouin zone is zero as shown in Fig.~\ref{fig:4band2}(d). Specifically, the normalized $x$- and $y$-components of the spin are ill-defined in the Brillouin zone as the unnormalized components are zero, corresponding to noise shown in Fig.~\ref{fig:Sconfig2}(b) and (c). The $z$-component of the ground state spin expectation value also goes to zero at certain $
\boldsymbol{k}$ points as shown in Fig.~\ref{fig:Sconfig2}(d), resulting in zero magnitude for the spin expectation value at these momenta.

\subsubsection{Toy models for chiral skyrmion superconductor phases}
As the generalized particle-hole operator $\mc{C}'$ is the product of the particle-hole operator $\mc{C}$ and the spatial inversion operator $\mc{I}$, we naturally expect the skymion phases in centrosymmetric superconductors. Indeed, this has already been confirmed for tight-binding models of Sr\textsubscript{2}RuO\textsubscript{4}~\cite{ueno2013}. To better understand when topological skyrmion phases are possible in such physical settings, we first consider a generic four-band model for a superconductor written in BdG formalism. The Hamiltonian $\mc{H}$ takes the form
\begin{equation}
\mc{H} = \sum_{\bk} \Psi^{\dagger}_{\bk} \mc{H}(\bk) \Psi_{\bk}/2,
\end{equation}
where
\begin{equation}
\mc{H}(\bk) = \begin{pmatrix}
\mc{E}(\bk) & \Delta(\bk) \\
\Delta^{\dagger}(\bk) & -\mc{E}^{T}(-\bk) \\
\end{pmatrix}
\end{equation}
and $\Psi_{\bk} = \left(c^{}_{\bk s}, c^{}_{\bk s'}, c^{\dagger}_{-\bk s}, c^{\dagger}_{-\bk s'} \right)^t$. Here, $c^{}_{\bk s}$ is the annihilation operator of electrons with momentum $\bk = \left( k_x, k_y \right)$ and spin angular momentum DOF $s \in \{ \uparrow, \downarrow \}$. $\mc{E}(\bk)$ is the Hamiltonian of the normal state, and $\Delta(\bk)$ is the gap function of the superconductor.

This model possesses $\mc{C}'$ symmetry when $-\epsilon^T(-\bk) = -\epsilon^*(\bk)$, which is satisfied for $\epsilon(\bk)$ even in momentum $\bk$. Therefore, to construct some representative example chiral topological skyrmion superconductor Hamiltonains, we take the normal state Hamiltonian to describe a Chern insulator that is even in momentum. One suitable example model is the following, from Sticlet~\emph{et al.}~\cite{Sticlet2012}:
\begin{equation}
    \epsilon(\bk) = \alpha\left[ \cos(k_x)\sigma_x + \cos(k_y)\sigma_y  \right] + \beta \cos(k_x+k_y)\sigma_z,
    \label{eq:Sticlet}
\end{equation}
which realizes topologically-distinct insulating phases characterized by Chern numbers of $0$, $\pm 2$ for the lower band.
We may then combine this with a pairing matrix $\Delta(\bk)$ describing either spin singlet pairing $\Delta_s(\bk)$ or spin-triplet pairing $\Delta_t(\bk)$. We take
\begin{align}
    \Delta_s(\bk) &= i \Delta_0 \sigma_y \label{eq:ssp}\\
    \Delta_t(\bk) &= i\Delta_0 \left(\boldsymbol{d}(\bk) \cdot \boldsymbol{\sigma}\right)\sigma_y,
    \label{eq:stp}
    \end{align}
    with the $\boldsymbol{d}$-vector of the spin-triplet pairing term here chosen specifically to be $\boldsymbol{d}(\bk) =\sin(k_y) \hat{x} - \sin(k_x)\hat{y}$ as one example. This $\boldsymbol{d}$-vector choice has previously been proposed as characterizing Sr$_2$RuO$_4$ in the high-field phase~\cite{ueno2013}.

\begin{figure}[t]
\includegraphics[width=0.5\textwidth]{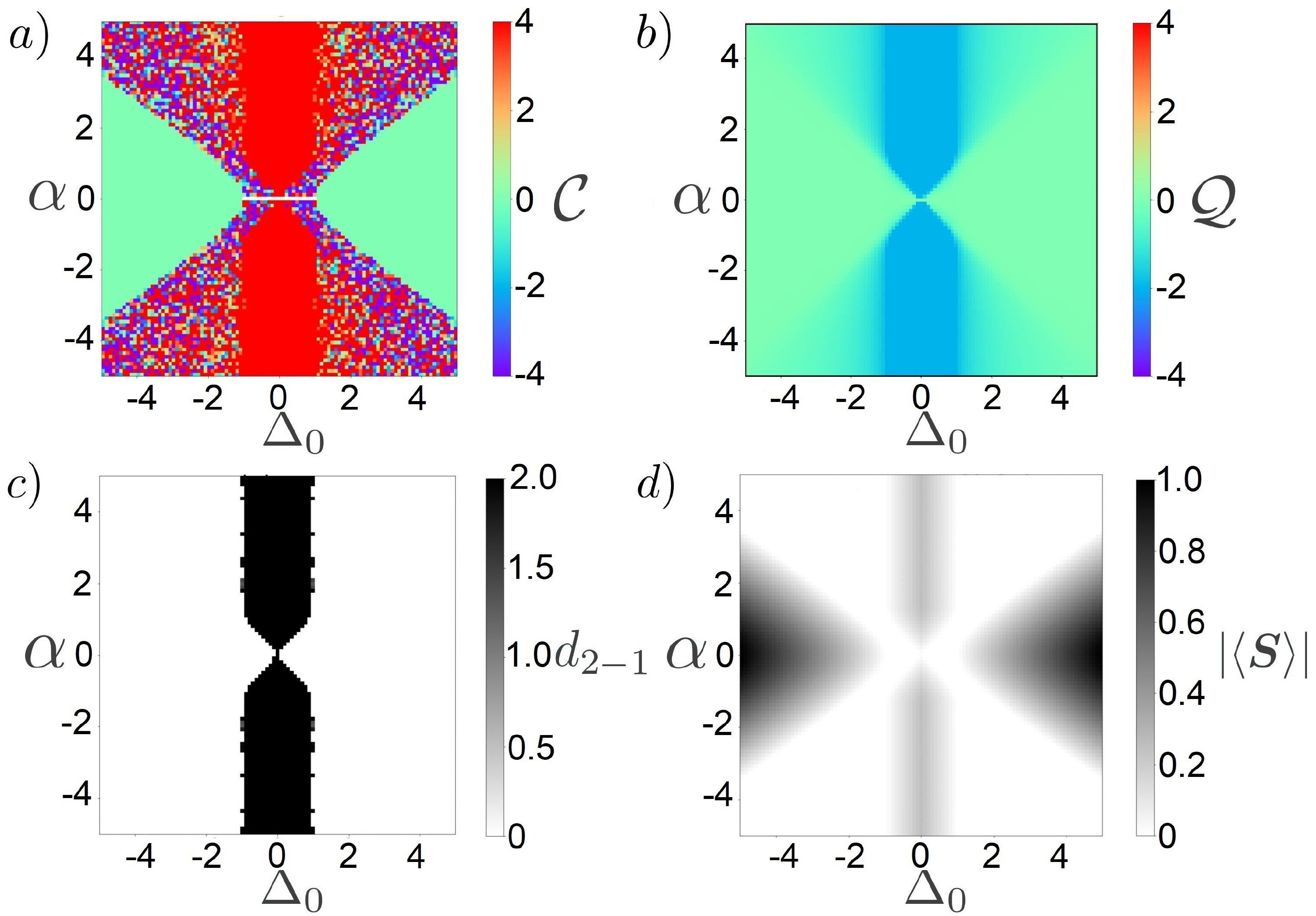}
\caption{Phase diagrams for BdG Hamiltonian with normal state Eq.~\eqref{eq:Sticlet}(with $\beta=1$)} and spin-singlet pairing given by gap function Eq.~\eqref{eq:ssp} at half-filling as a function of pairing strength $\Delta_0$ and hopping integral $\alpha$ for the (a) total Chern number $\mc{C}$, (b) skyrmion number  $\mc{Q}$, (c) minimum direct superconducting gap $d_{2-1}$, and (d) minimum magnitude of the ground-state spin expectation value $|\langle S \rangle |$.
\label{fig: PD ssp}
\end{figure}

We may now compute phase diagrams for these two models characterizing spin-singlet and spin-triplet superconductivity, respectively. We first consider the case of spin-singlet pairing. To characterize the topology of the model, we compute the total Chern number $\mc{C}$ and skyrmion number $\mc{Q}$ assuming two bands are filled, as well as the minimum direct gap between the third-lowest and second-lowest bands, $d_{2-1}$, and the minimum magnitude of the ground state spin expectation value over the Brillouin zone, $|\langle \boldsymbol{S} \rangle |$. Each quantity is computed as a function of parameter $\alpha$ controlling some of the kinetic terms and the pairing strength $\Delta_0$. These results for the case of spin-singlet pairing are shown in  Fig.~\ref{fig: PD ssp}.

The total Chern number $\mc{C}$ phase diagram shown in Fig.~\ref{fig: PD ssp} (a) may be divided into three distinct parts: a topologically stable region with total Chern number of $4$ shown in red, which has an hourglass shape, two triangular regions where the Chern number is well-quantized at zero shown in green, and regions where the total Chern number is highly unstable. In comparison, the phase diagram for the skyrmion number in Fig.~\ref{fig: PD ssp} (b) possesses a tan hourglass region with $\mc{Q}=2$ where the total Chern number $\mc{C}=4$. This is further evidence that the skyrmion number differs from the total Chern number by a factor of $2$ at the four band level when they are each topologically stable. There are then two triangular regions with the skyrmion number very well-quantized at zero, and regions around the hourglass where the skyrmion number appears unquantized but close to zero.
\begin{figure}[t]
\includegraphics[width=0.5\textwidth]{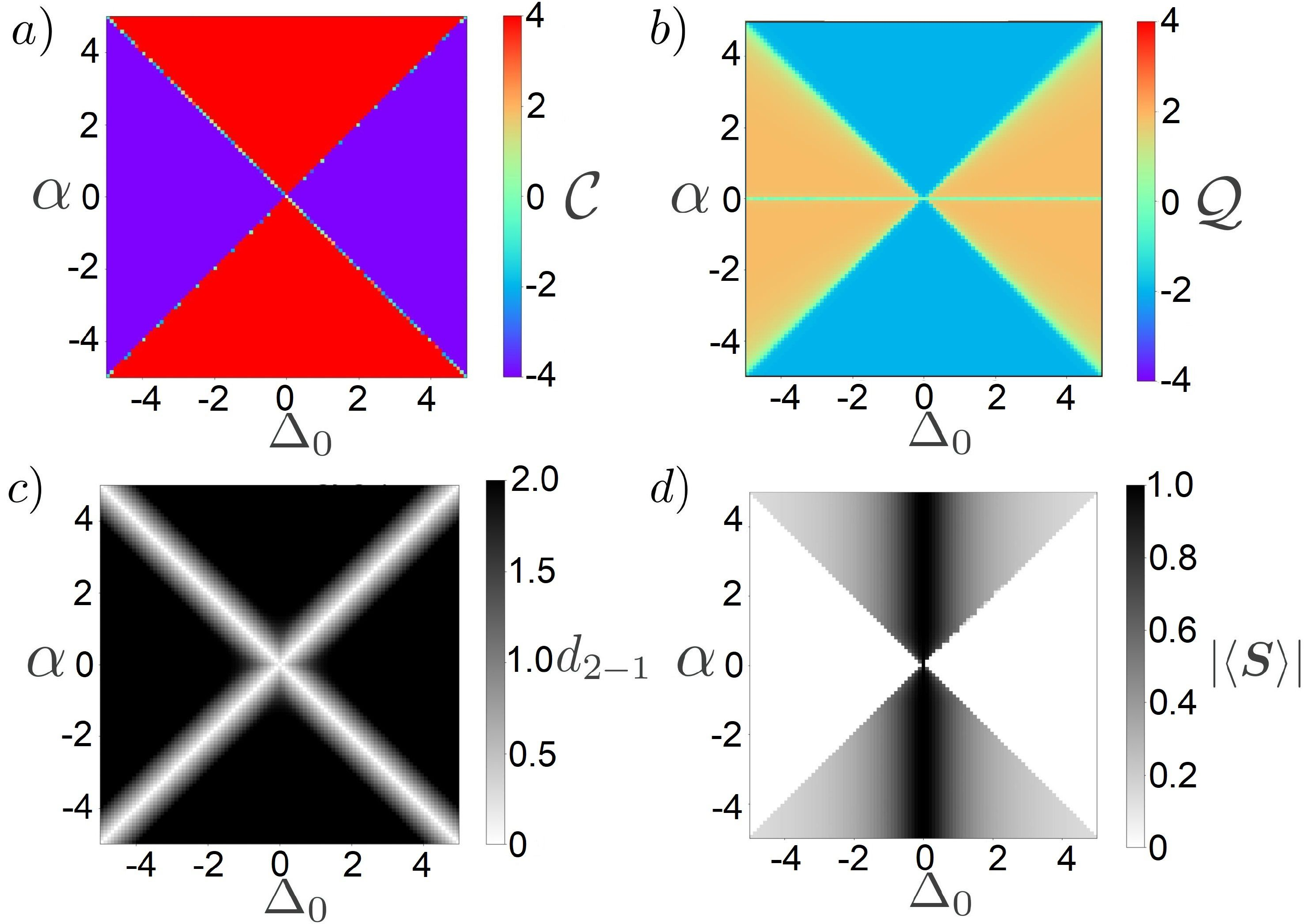}
\caption{Phase diagrams for BdG Hamiltonian with normal state Eq.~\eqref{eq:Sticlet}(with $\beta=1$)} and spin-triplet pairing given by gap function Eq.~\eqref{eq:stp} at half-filling as a function of pairing strength $\Delta_0$ and hopping integral $\alpha$ for the (a) total Chern number $\mc{C}$, (b) skyrmion number  $\mc{Q}$, (c) minimum direct superconducting gap $d_{2-1}$, and (d) minimum magnitude of the ground-state spin expectation value $|\langle S \rangle |$.
\label{fig: PD stp}
\end{figure}

To better understand the model, we also examine the phase diagram for the minimum direct gap $d_{2-1}$, shown in Fig.~\ref{fig: PD ssp} (c). We see that it is finite only for the hourglass region, explaining the instability observed in Fig.~\ref{fig: PD ssp} (a) and (b). In addition, we compute the phase diagram for the minimum magnitude of the ground state spin expectation value, $|\langle \boldsymbol{S} \rangle |$. We see that it is finite in the hourglass region as required for a topologically-stable value of the skyrmion number, but zero in the region where the skyrmion number is not well-quantized. This is as expected, since the skyrmion number is computed with the normalized ground state spin expectation value, and is therefore unstable when the spin is not normalizable somewhere in the Brillouin zone.

We now consider the case of spin-triplet pairing. By tuning the parameters $\alpha$ and $\Delta_0$, we again compute the total Chern number $\mc{C}$ and skyrmion number $\mc{Q}$ assuming half-filling of bands, as well as the minimum direct gap over the Brillouin zone $d_{2-1}$ and the minimum ground state spin expectation value over the Brillouin zone $| \langle \boldsymbol{S}\rangle |$.

The total Chern number phase diagram, shown in  Fig.~\ref{fig: PD stp} (a), is topologically-stable almost everywhere in the phase diagram at values $\mc{C}=\pm4$. The skyrmion number phase diagram, shown in  Fig.~\ref{fig: PD stp} (b), is similar in form, yet the tan regions with $\mc{Q}=2$ appear far more stable than the blue regions with $\mc{Q}\approx -2$. We see, again, that topological phase transitions occur where the minimum direct gap over the Brillouin zone between the two middle bands, $d_{2-1}$, goes to zero, as shown in Fig.~\ref{fig: PD stp} (c). We gain insight into the difference between the $\mc{Q}=2$ regions and the $\mc{Q} = -2$ regions by examining the phase diagram for the minimum ground state spin expectation value over the Brillouin zone, $|\langle \boldsymbol{S} \rangle |$: regions with $\mc{Q}=2$ have finite spin everywhere in the BZ, while $|\langle \boldsymbol{S} \rangle |$ is zero in magnitude somewhere the BZ for $\mc{Q}=-2$ regions. Thus these regions are topologically unstable according to the skyrmion number while being topologically stable according to the total Chern number, further indicating that these two topological invariants are decoupled from one another at the four band level. Remarkably, however, the ground state spin texture nearly forms skyrmions even in these cases where the spin is somewhere zero in magnitude in the BZ, similar to results for the mirror subsectors of the previously-considered tight-binding model for Sr$_2$RuO$_4$~\cite{ueno2013}.
\begin{figure}[t]
\centering
\includegraphics[width=0.5\textwidth]{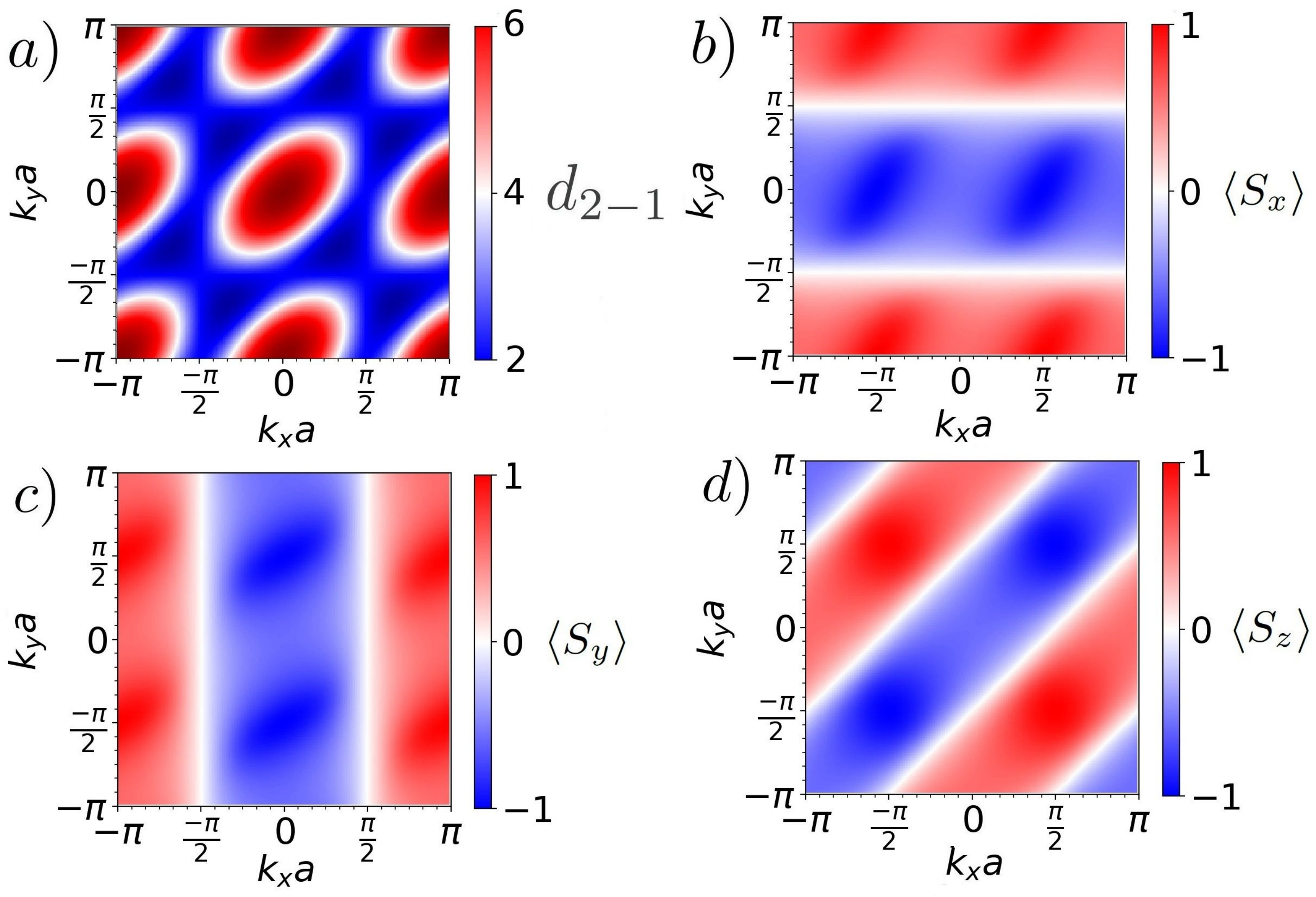}
\caption{Characterization of a representative point in the phase diagrams for BdG Hamiltonian with normal state Hamiltonian Eq.~\eqref{eq:Sticlet} and spin-singlet gap function Eq.~\eqref{eq:ssp} shown in Fig.~\ref{fig: PD ssp} ($\Delta_0=1$ and $\alpha=2$). Minimum direct superconducting gap $d_{2-1}$ over the first Brillouin zone is shown in (a). $x$, $y$, and $z$ components of the normalized ground-state spin expectation value, $\langle S_x \rangle$, $\langle S_y \rangle$, and $\langle S_z \rangle$, over the first Brillouin zone are shown in (b), (c), and (d), respectively.}
\label{fig: ssp texture}
\end{figure}

Noticeably, only type-I topological phase transitions occur in each model, that is, only topological phase transitions that occur with a closing of the direct gap between the two middle bands. This results from the four-band $\mc{C}'$-invariant Hamiltonians lacking a suitable DOF, such as orbital angular momentum, into which spin angular momentum can flow. Although one could interpret the Bloch Hamiltonian as corresponding to a system with a spin half DOF and a two-fold orbital DOF, the $\mc{C}'$ invariance of the model restricts the orbital DOF such that only the type-I topological phase transition occurs at the four-band level. In Hamiltonians with six or more bands, however, topology of an orbital DOF is no longer ensured by $\mc{C}'$ symmetry as discussed in work introducing the topological skyrmion phases of matter and observed for tight-binding models describing superconducting Sr\textsubscript{2}RuO\textsubscript{4}~\cite{ueno2013}, making the type-II topological phase transition possible.

\begin{figure}[t]
\includegraphics[width=0.5\textwidth]{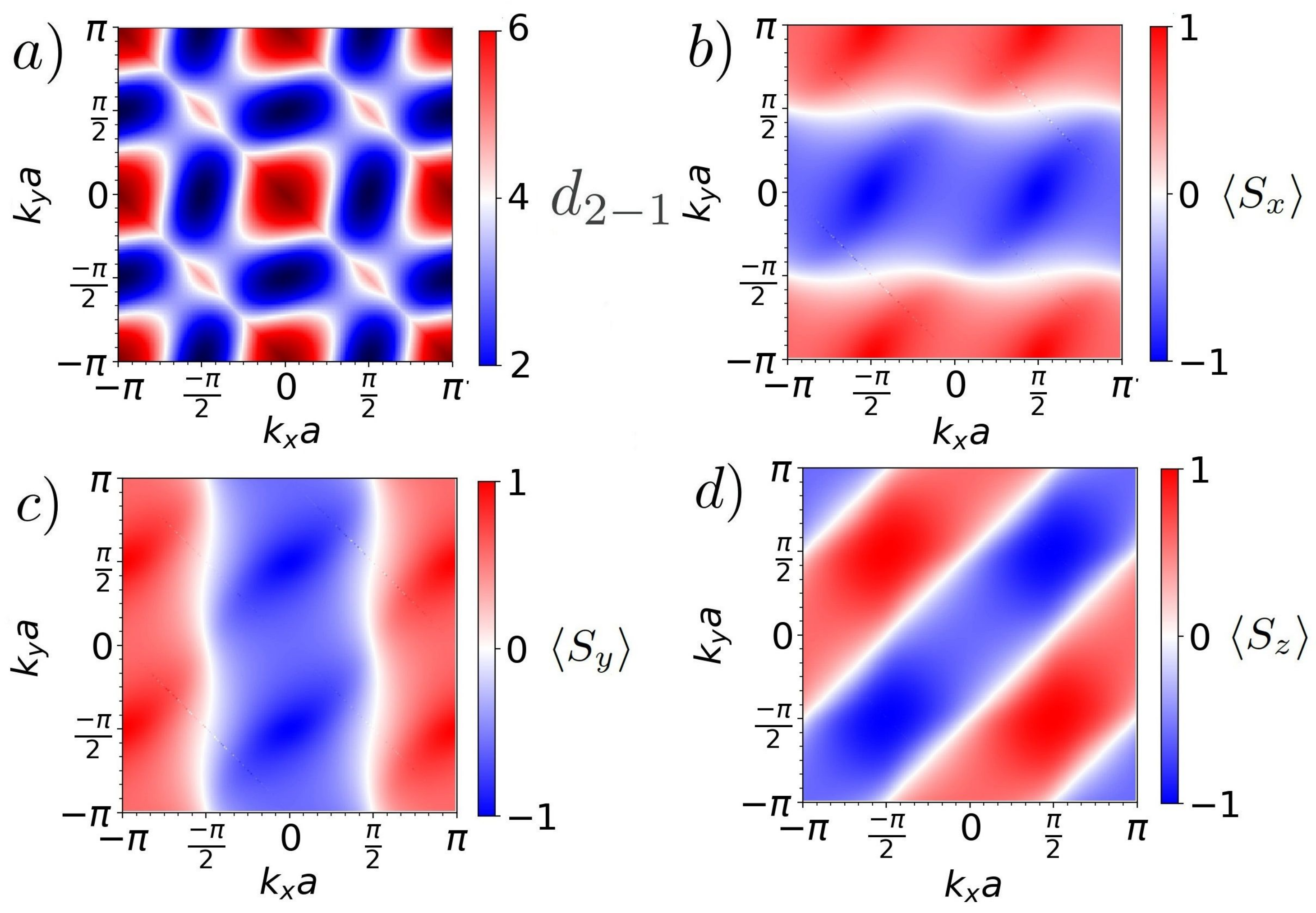}
\caption{ Characterization of a representative point in the phase diagrams for BdG Hamiltonian with normal state Hamiltonian Eq.~\eqref{eq:Sticlet} and spin-singlet gap function Eq.~\eqref{eq:stp} shown in Fig.~\ref{fig: PD stp} ($\Delta_0=1$ and $\alpha=2$). Minimum direct superconducting gap $d_{2-1}$ over the first Brillouin zone is shown in (a). $x$, $y$, and $z$ components of the normalized ground-state spin expectation value, $\langle S_x \rangle$, $\langle S_y \rangle$, and $\langle S_z \rangle$, over the first Brillouin zone are shown in (b), (c), and (d), respectively.}
\label{fig: stp texture}
\end{figure}

We fix a particular choice of parameters $\Delta_0=1, \alpha=2$ to study the spin textures in the $\mc{Q}=-2$ skyrmion phase. As shown in Fig.~\ref{fig: ssp texture} and Fig.~\ref{fig: stp texture}, the system is always gapped and each component of the spin expectation is always well-defined and varies continuously in the BZ. In comparison with the $\mc{Q}=-1$ case in Fig.~\ref{fig:Sconfig1}, the $x$- and $y$-component of the spin expectation demonstrate very similar behaviour, while the $z$-component winds once around the BZ for $\mc{Q}=-1$ and twice for $\mc{Q}=-2$, demonstrating the nature of $\mc{Q}$ as a winding number.

\section{Three-dimensional chiral topological skyrmion phase}\label{3}

\subsection{Section overview}

In this second section of the paper, we introduce the three-dimensional counterpart of the two-dimensional chiral topological skyrmion phase, taking advantage of lessons gained in constructing suitable models in two-dimensions, which are discussed in Section II.

We first explain why such a three-dimensional counterpart phase is anticipated, before explaining how toy models for the 3D phase are constructed taking inspiration from the topological phase known as the Hopf insulator~\cite{moore2008}. To do this, we explain the defining characteristic of the 3D chiral skyrmion phase and how analogous physics defines the Hopf insulator phase, before utilizing techniques from Section I to construct a toy model Hamiltonian for the 3D chiral topological skyrmion phase from a two-band Hamiltonian for the Hopf insulator and a two-band Hamiltonian for its $\mc{C}'$ partner.

We then introduce the topological invariant for characterization of the 3D chiral skyrmion phase, and discuss its numerical computation. We then use this invariant to characterize some canonical toy models for the 3D chiral skyrmion phase. Finally, motivated by past work on realization of unpaired Majorana zero-modes in two-dimensional systems with a defect that are characterized by the Hopf insulator Hamiltonian, we show that the Hamiltonian for the 3D chiral skyrmion phase with the same boundary conditions also realizes topologically-protected zero-modes.

\subsection{Construction of models for the three-dimensional chiral skyrmion phase}

While a three-dimensional counterpart of the chiral topological skyrmion phase is expected in Hamiltonians with $\mc{C}'$ symmetry, as they correspond to the non-trivial homotopy group $\pi_3\left(\mathrm{Sp}(2N) /\mathrm{U}(N) \right)  = \mathbb{Z}_2$, it is not immediately clear how to construct a model realizing this topological phase.

We expect the three-dimensional phase to correspond to formation of a hedgehog in the ground state spin expectation value over the three-dimensional Brillouin zone, rather than a baby skyrmion as in the two-dimensional case. Given that the atomic model for the two-dimensional chiral skyrmion phase is Eq.~\ref{atomicham}, which may be constructed from a two-band Chern insulator and its $\mc{C}'$ partner, we can see that the atomic model for the three-dimensional case would be constructed from some two-band Hamiltonian that realizes a momentum-space hedgehog and its $\mc{C}'$ partner. There is such a two-band model: Hamiltonians realizing the Hopf insulator~\cite{moore2008} phase of matter realize such hedgehogs in the spin or pseudospin texture over the Brillouin zone when the topological invariant characterizing the phase, the Hopf invariant, is non-trivial.
\begin{figure}[t]
\includegraphics[width=0.5\textwidth]{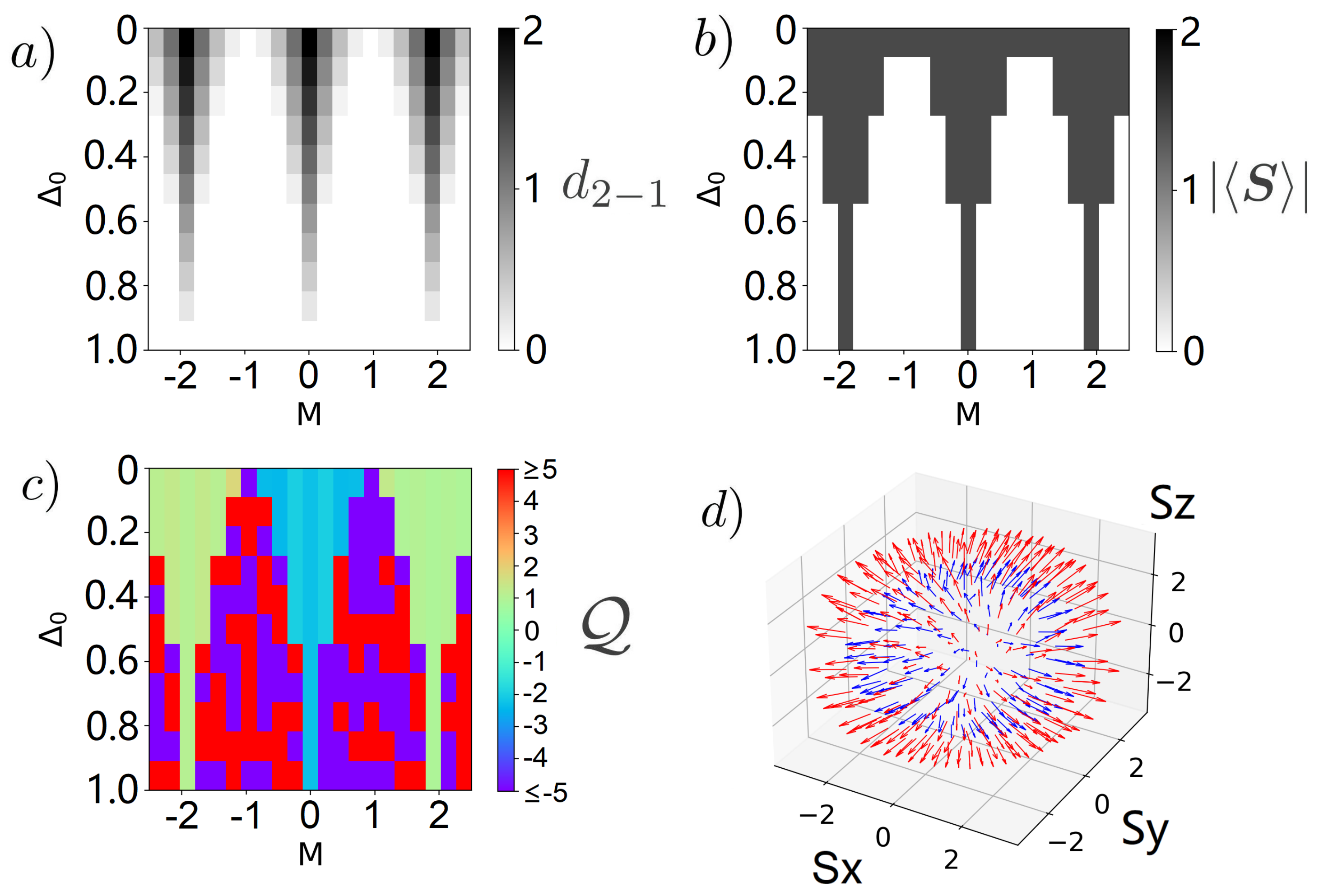}
\caption{Phase diagrams as a function of $M$ and $\Delta_0$ for the (a) minimum direct energy gap between the second- and third-lowest bands in energy, $d_{2-1}$, (b) minimum spin magnitude $|\langle S \rangle |$ over the Brillouin zone, and (c) skyrmion number $\mc{Q}$. (d) shows the hedgehog shape formed by the ground-state spin expectation value texture over the Brillouin zone for Hamiltonian Eq.~\eqref{eq:3D} plus spin singlet pairing Eq.~\eqref{eq:ssp} at half-filling in a plot of the ground state spin texture for two spherical shells of momenta centred at the origin in the 3D Brillouin zone with $|\bk|=\pi$ (red arrows) and $|\bk|=0.75\pi$ (blue arrows)}.
\label{fig:ssp3D}
\end{figure}
The two-band Bloch Hamiltonian $h(\bk)$ for the Hopf insulator takes the following form:
\begin{equation}
     h(\bk) = \sum_{i=1,2,3}d_i(\bk)\tau_i,
     \label{Hopf 2band}
\end{equation}
where
\begin{equation}
    d_i = z^\dagger \tau_i z,
\end{equation}
for which we take $z=(z_1,z_2)^T$ where
\begin{align}
    z_1 &= \sin k_x + i\sin k_y,\\
    z_2 &= \sin k_z + i(\cos k_x+\cos k_y+\cos k_z-M).
    \label{eq:M}
\end{align}
The four-band atomic model for the three-dimensional topological skyrmion phase of matter may then be written as
\begin{equation}
    \mc{H}(\bk) = \begin{pmatrix}
             h(\bk) & 0 \\
             0 & -h^{*}(\bk)
            \end{pmatrix}.
\label{eq:3D}
\end{equation}
As in two-dimensions, we can add additional, symmetry-allowed terms to this Bloch Hamiltonian, such as spin singlet pairing Eq.~\eqref{eq:ssp} or spin triplet pairing Eq.~\eqref{eq:stp}.
Phase diagrams are computed for these Hamiltonians with $M$ in Eq.~\eqref{eq:M} and $\Delta_0$ in Eq.~\eqref{eq:stp} as varying parameters, and the results are shown in Fig.~\ref{fig:ssp3D} and \ref{fig:stp3D}, respectively. We show that the system realizes a skyrmion phase for a range of parameters $\Delta_0$ and $M$ as characterized by the skyrmion number.

\subsection{Topological invariant in three-dimensions}

Here, we introduce the topological invariant of the three dimensional topological skyrmion phases of matter. To provide context, we first review the Hopf invariant. Although there are some similarities between the form of the Hopf invariant and the topological invariant of the 3D topological skyrmion phase, they are computed with different physical quantities and are therefore in general distinct.

The Hopf invariant is the topological charge of a vector field defined over the 3D Brillouin zone, where the vector is the projector onto occupied states. Following the paper introducing the Hopf insulator protected by $\mc{C}'$ symmetry in systems with $2$ or more bands~\cite{liu2017}, we consider a Hamiltonian $\mc{H}(\boldsymbol{k})$ which takes the following form when diagonalized,
\begin{equation}
    \mc{H}(\boldsymbol{k}) = \mathrm{diag}\left( \lambda_1(\boldsymbol{k}), -\lambda_1(\boldsymbol{k}), ..., \lambda_N(\boldsymbol{k}), -\lambda_N(\boldsymbol{k}) \right).
\end{equation}
Continuously deforming all of the positive eigenvalues to $+1$, and all of the negative eigenvalues to $-1$, the deformed Hamiltonian takes the form,
\begin{equation}
    \mc{H}(\boldsymbol{k}) = U^{}(\boldsymbol{k})  I^{}_{N,N} U^{\dagger}(\boldsymbol{k}),
\end{equation}
where $U(\boldsymbol{k})$ is the matrix of eigenvectors and $I^{}_{N,N} = I^{}_N \otimes \sigma_z$, where $I^{}_N$ is the $N\times N$ identity matrix and $\sigma_z$ is the Pauli matrix $\sigma_z = \mathrm{diag}(1,-1)$. We can then define the Hopf invariant in terms of the projectors onto occupied states as in Liu~\emph{et al.}~\cite{liu2017} as
\begin{equation}
    n_{\omega} = {1 \over 24 \pi ^2} \int d^3 k \left(U^{-1} dU \right)^3,
\end{equation}
and therefore denote it as a \textit{projector topological invariant}. If the only degree of freedom labeling the fermionic creation and annihilation operators in a given $\boldsymbol{k}$-sector is spin, this projector also corresponds to the ground-state spin expectation value. If the operators are labeled not just by spin, but also by other degrees of freedom, such as an orbital degree of freedom, or a particle-hole degree of freedom, the projector onto occupied states becomes distinct from the ground-state spin expectation value vector. In this case, one can still compute the topological charge for the projector vector field, but also a second topological charge, for the spin expectation value vector. In some cases, this spin topological charge quantizes and serves as a second topological invariant distinct from the projector-based topological invariant, which we define as a \textit{spin topological invariant}. The 3D topological skyrmion phase corresponds to the system realizing non-trivial, quantized spin topological charge, when the spin expectation value vector is distinct from the projector onto occupied states.

For these cases where the projector topological invariant and spin topological invariant are computed with distinct physical quantities, we may then write the topological invariant of the 3D chiral skyrmion phase as
\begin{align}
    \mc{Q}_{3D} = -{1 \over 8 \pi} \int_{BZ}d^3k \boldsymbol{A}_{\mc{Q}} \cdot \boldsymbol{\Omega}_{\mc{Q}}
    \label{3Dinvar}
\end{align}
where here $\boldsymbol{A}_{\mc{Q}}$ is the skyrmion connection defined in Eq.~\ref{skyrmconnect}, and $\boldsymbol{\Omega}_{\mc{Q}}$ is the skyrmion curvature defined in Eq.~\ref{skyrmcrv}.

In this work, we restrict ourselves to four band models constructed from a two-band Hamiltonian and its $\mc{C}'$ partner, which each realize a topological phase characterized by the Hopf invariant. We may therefore compute a Hopf invariant using projectors onto occupied states as defined above, but also a separate 3D skyrmion number.

The 3D skyrmion number $\mc{Q}_{3D}$ may therefore be computed using Eq.~\ref{skyrmcrvpart} for the skyrmion curvature, and Eq.~\ref{skyrmconnpart} for the skyrmion connection, computing the derivatives numerically. Using this invariant $\mc{Q}_{3D}$, we characterize the topology of the toy models presented for the 3D chiral skyrmion phase.

\begin{figure}[t]
\includegraphics[width=0.5\textwidth]{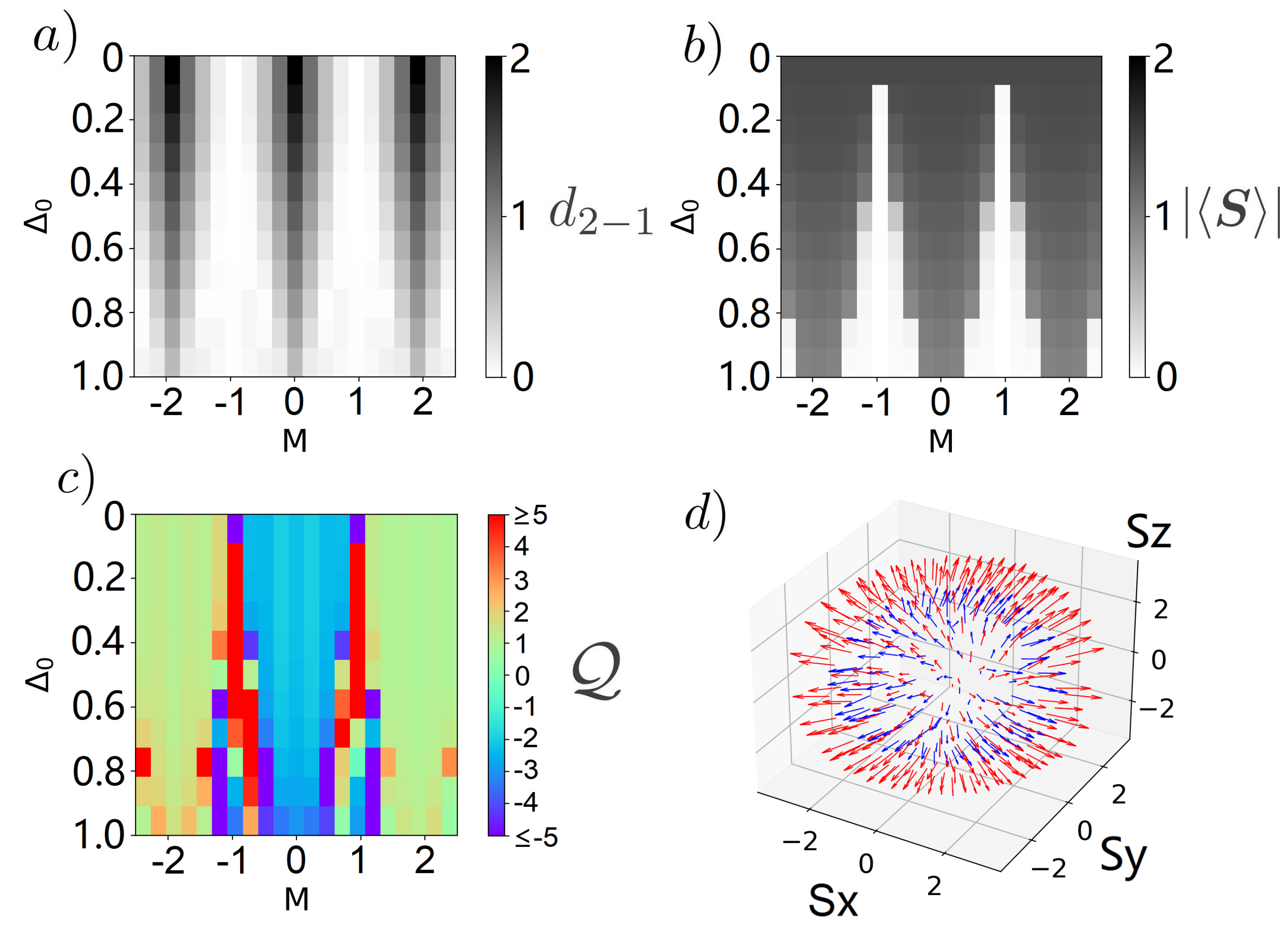}
\caption{Phase diagrams as a function of $M$ and $\Delta_0$ for the (a) minimum direct energy gap between the second- and third-lowest bands in energy, $d_{2-1}$, (b) minimum spin magnitude $|\langle S \rangle |$ over the Brillouin zone, and (c) skyrmion number $\mc{Q}$. (d) shows the hedgehog shape formed by the ground-state spin expectation value texture over the Brillouin zone for Hamiltonian Eq.~\eqref{eq:3D} plus spin triplet pairing Eq.~\eqref{eq:stp} at half-filling in a plot of the ground state spin texture for two spherical shells of momenta centred at the origin in the 3D Brillouin zone with $|\bk|=\pi$ (red arrows) and $|\bk|=0.75\pi$ (blue arrows)}.
\label{fig:stp3D}
\end{figure}

A similar set of phase diagrams is computed for the Hamiltonian in Eq.~\eqref{eq:3D} with free parameter $M$ \textit{plus} additional spin triplet pairing term Eq.~\eqref{eq:stp} with free parameter $\Delta_0$. The results are shown in Fig.~\ref{fig:ssp3D} and \ref{fig:stp3D}. We show that the system can be in the three-dimensional topological skyrmion phase for a range of parameters $\Delta_0$ and $M$ as characterized by the skyrmion number.

\subsection{Bulk-boundary correspondence}

We now explore the bulk-boundary correspondence associated with non-trivial $\mc{Q}_{3D}$. Bulk-boundary correspondence is expected when a projector topological invariant is non-trivial, and when symmetries protecting the topological phase in the bulk are respected in the bulk and by open boundary conditions~\cite{schnyder2008, Ryu_2010}. Here, we instead investigate whether the bulk-boundary correspondence is realized for a topological phase characterized by trivial projector topological invariant, the Hopf invariant $n_{\omega}$, and non-trivial spin topological invariant $\mc{Q}_{3D}$.

To provide context, we first review the bulk-boundary correspondence of the Hopf insulator, as there are at least two different kinds of bulk-boundary correspondence, which have been considered for this system and are relevant to the present work. These two types of bulk-boundary correspondence are distinguished by different open boundary conditions. The first kind of bulk-boundary correspondence of the Hopf insulator is discussed in Moore~\emph{et al.}~\cite{moore2008} introducing the Hopf insulator phase. This corresponds to open boundary conditions in the $\hat{z}$-direction, and periodic boundary conditions in the $\hat{x}$ and $\hat{y}$ directions. Under these boundary conditions, non-trivial Hopf invariant corresponds to gapless boundary modes known as Fermi rings localized on the top and bottom surfaces.

Here, it is important to distinguish between the Hopf insulator phase realized in a two-band system, and the Hopf insulator realized in systems with more than two bands protected by generalized particle-hole symmetry $\mc{C}'$~\cite{LiuHopf2017}. The two-band Hopf insulator has $\mathbb{Z}$ topological classification and Fermi ring surface states due to bulk-boundary correspondence. The $\mc{C}'$-protected Hopf insulator with more than two bands instead has $\mathbb{Z}_2$ topological classification. Furthermore, as the $\mc{C}'$ operation involves particle-hole conjugation in combination with spatial inversion, opening boundary conditions in the $\hat{z}$-direction generically breaks $\mc{C}'$ near the surfaces. Gapless surface states are then not topologically-protected. As our 3D chiral skyrmion phase is protected by $\mc{C}'$ symmetry, we do not expect to find this kind of bulk-boundary correspondence.

A second kind of bulk-boundary correspondence has also been considered in the context of the Hopf insulator~\cite{yan2017}. In this case, boundary conditions are first opened in the $\hat{x}$- and $\hat{y}$-directions, while periodic boundary conditions are retained in the $\hat{z}$-direction. The Hamiltonian is therefore defined in terms of two real-space coordinates $x$ and $y$, and one momentum component $k_z$. Then, one performs the substitution $k_z \rightarrow \theta (x,y)$. This spatially-varying $\theta$ is interpreted as an angle in the $x-y$ plane, which can characterize a zero-dimensional defect.

\begin{figure}[t]
\includegraphics[width=0.5\textwidth]{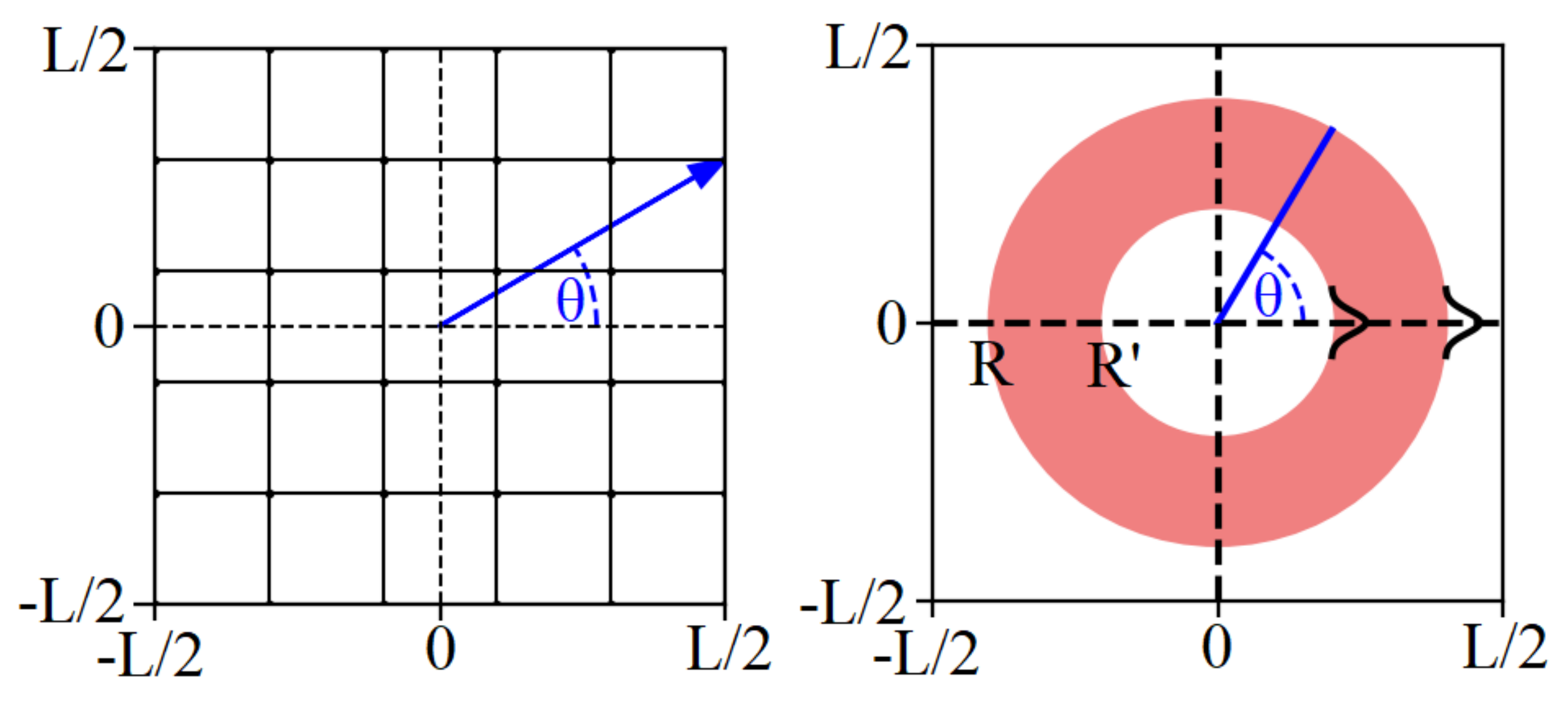}
\caption{(a) The two-dimensional lattice with the defect located at $(0,0)$ and $\theta=\arctan(y/x)$. (b) Schematic illustration of the annulus geometry considered for low-energy analytical theory of the defect bulk-boundary correspondence. For the Hopf insulator, at $\theta=0$, one unpaired Majorana zero-mode is localized on the inner boundary and one on the outer boundary of the annulus, respectively, as shown in this schematic.}
\label{fig:annulus}
\end{figure}

We first review the results of Yan~\emph{et al}~\cite{yanzhongbo2017} on this second kind of bulk-boundary correspondence, in particular discussing the nature of the gapless boundary modes. The boundary conditions are shown in Fig.~\ref{fig:annulus} (a) for the Hopf insulator lattice model.  A corresponding annulus geometry for the low-energy theory is shown in Fig.~\ref{fig:annulus} (b). This geometry yields a bulk-boundary correspondence when the Hopf invariant is non-trivial, and the Hopf insulator Bloch Hamiltonian realizes a hedgehog in its (pseudo)spin texture centered at the origin of the three-dimensional Brillouin zone. According to the topological classification of defects by Teo and Kane~\cite{teo2010}, the defect can correspond to a non-trivial homotopy group and realize a topologically non-trivial, odd-parity state. As the defect is a zero-dimensional system within the Brillouin zone, it is, however, not immediately clear how to realize unpaired Majorana zero-modes from this odd-parity state. A key insight of Yan~\emph{et al.}~\cite{yanzhongbo2017} was that re-interpreting the Hopf insulator Bloch Hamiltonian as describing a two-dimensional system with a defect means the zero-dimensional defect of the three-dimensional Brillouin zone corresponds to an effectively one-dimensional system along the line $\theta=0$ in the 2D system with a defect. Then, similarly to one-dimensional wire proposals for realizing unpaired Majorana zero-modes, an unpaired Majorana zero-mode is realized at each end of this $\theta=0$ wire.

\begin{figure}[t]
\includegraphics[width=0.5\textwidth]{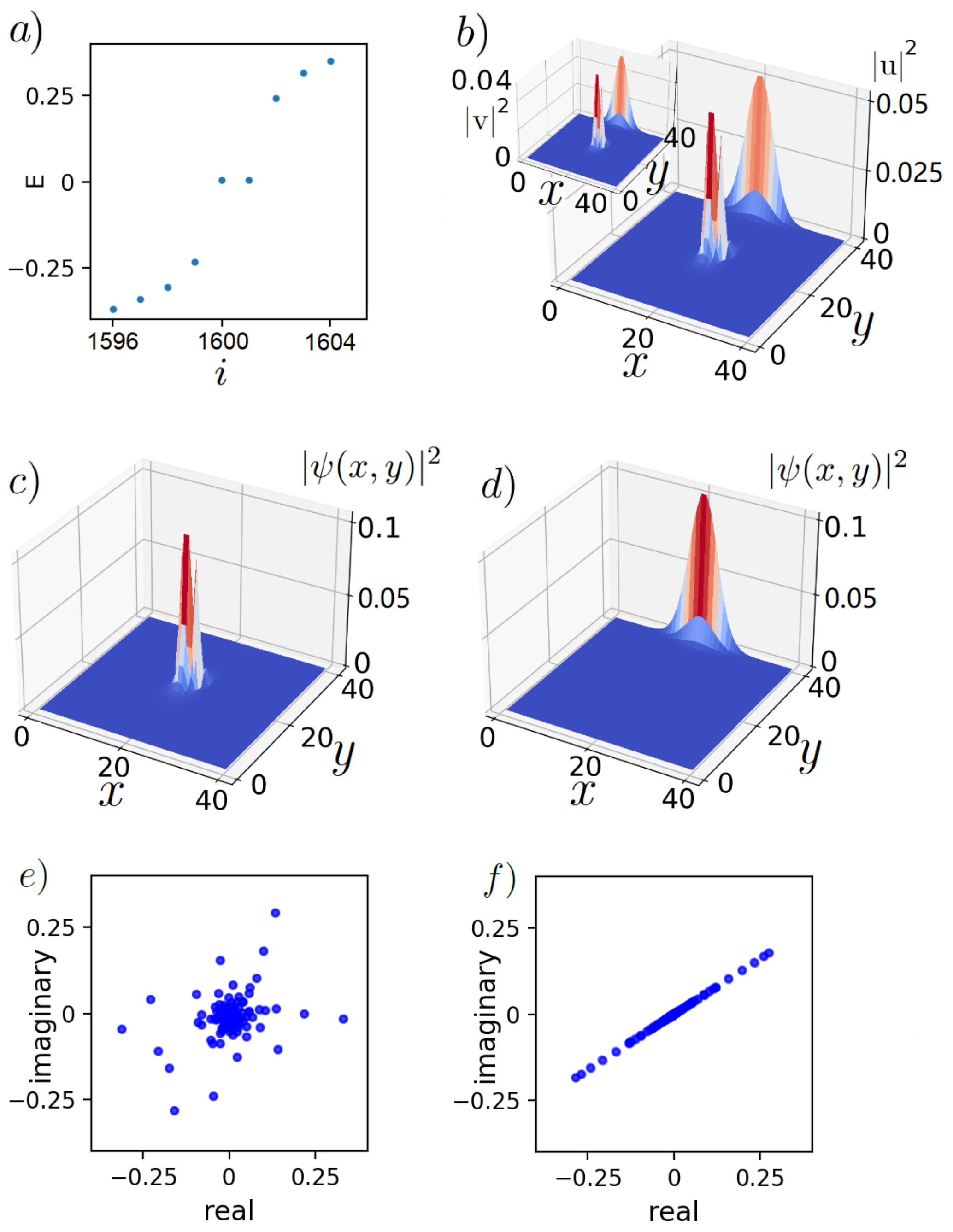}
\caption{For the Hopf Hamiltonian given by Eq.~\eqref{Hopf 2band} with open boundary conditions in the $x$ and $y$ directions and the $z$ component of momentum $k_z$ replaced with $\theta$ characterizing a defect, (a) spectrum vs. eigenvalue index $i$ in the vicinity of zero-energy, showing a two-fold degeneracy at zero-energy. (b) is the probability density plot of particle component $u$ with the hole component $v$ in the inset of one of the two zero-energy states outputted by numerical diagonalization. (c) and (d) show the strongly localized unpaired Majorana zero-modes computed as linear combinations of the two zero-energy states yielded by numerical diagonalization. (e) and (f) show each complex number in the eigenvector for each of the unpaired Majorana bounds plotted in the complex plane.}
\label{fig: Hopf2band}
\end{figure}

For such a finite-size, two-dimensional system described by the Hopf insulator Hamiltonian with $k_z$ replaced by angle $\theta$ characterizing a defect at the center of the system, we first numerically compute the spectrum and eigenstates of the Hopf insulator Bloch Hamiltonian as is done in Yan~\emph{et al}~\cite{yanzhongbo2017}, with a defect located at $(0,0)$ and $\theta=\arctan(y/x)$. The results are shown in Fig.~\ref{fig: Hopf2band} for a finite-size system with $40\times40$ lattice sites. These results include the low-energy spectrum, the particle and hole component probability densities of a zero-energy eigenstate, and the probability densities for each unpaired Majorana zero-mode,  shown in Fig.~\ref{fig: Hopf2band} (a), (b), (c), and (d), respectively. As expected, the unpaired Majorana zero-mode at the boundary occurs for $\theta=0$, corresponding to the location of the zero-dimensional defect in the three-dimensional Brillouin zone of the Hopf insulator.

We then additionally characterize the resultant unpaired Majorana zero-modes in greater detail. Specifically, after taking linear combinations of the two lowest energy eigenstates of the Hopf insulator Hamiltonian with these boundary conditions to compute arrays with $40\times 40\times 2$ entries representing the unpaired Majorana zero-mode state wavefunction, we then plot each of these entries in the complex plane as a point, to better understand later results for the 3D chiral skyrmion phase. These representations in the complex plane of the unpaired Majorana bound states localized at the defect and boundary are shown in Fig.~\ref{fig: Hopf2band} (e) and (f), respectively. The MZM at the boundary appears as a straight line in the complex plane. This is expected for an unpaired MZM, which has a purely real wavefunction up to some phase degree of freedom. The structure of the MZM at the defect, in contrast, is polluted. This complicated structure of the zero-mode state localized at the defect is unexpected for an unpaired MZM: a key feature of unpaired MZMs is that they should be straight lines i.e. they are purely real up to a phase degree of freedom\cite{alicea2012, read2000}. The unpaired MZM localized at the defect likely loses this feature because of coupling to higher energy states at $\theta \neq 0$ due to the small radius of the defect.

Given that the 3D chiral skyrmion phase toy model presented here is constructed from the Hopf insulator Hamiltonian, we anticipate a similar bulk-boundary correspondence for the 3D chiral skyrmion phase, when describing a two-dimensional system with open boundary conditions and a defect, as occurs for the Hopf insulator Hamiltonian itself. We consider the following 3D chiral skyrmion phase Hamiltonian,
\begin{align}
    H_{\textnormal{3D}}(\bk)=\mc{H}(\bk)+\Delta_t(\bk)
    \label{eq:Hopf 3D}
\end{align}
where $\mc{H}(\bk)$ is the 4-band Hopf Hamiltonian in Eq.~\eqref{eq:3D} and $\Delta_t(\bk)$ is the spin triplet pairing matrix in Eq.~\eqref{eq:stp}.

\begin{figure}[t]
\includegraphics[width=0.5\textwidth]{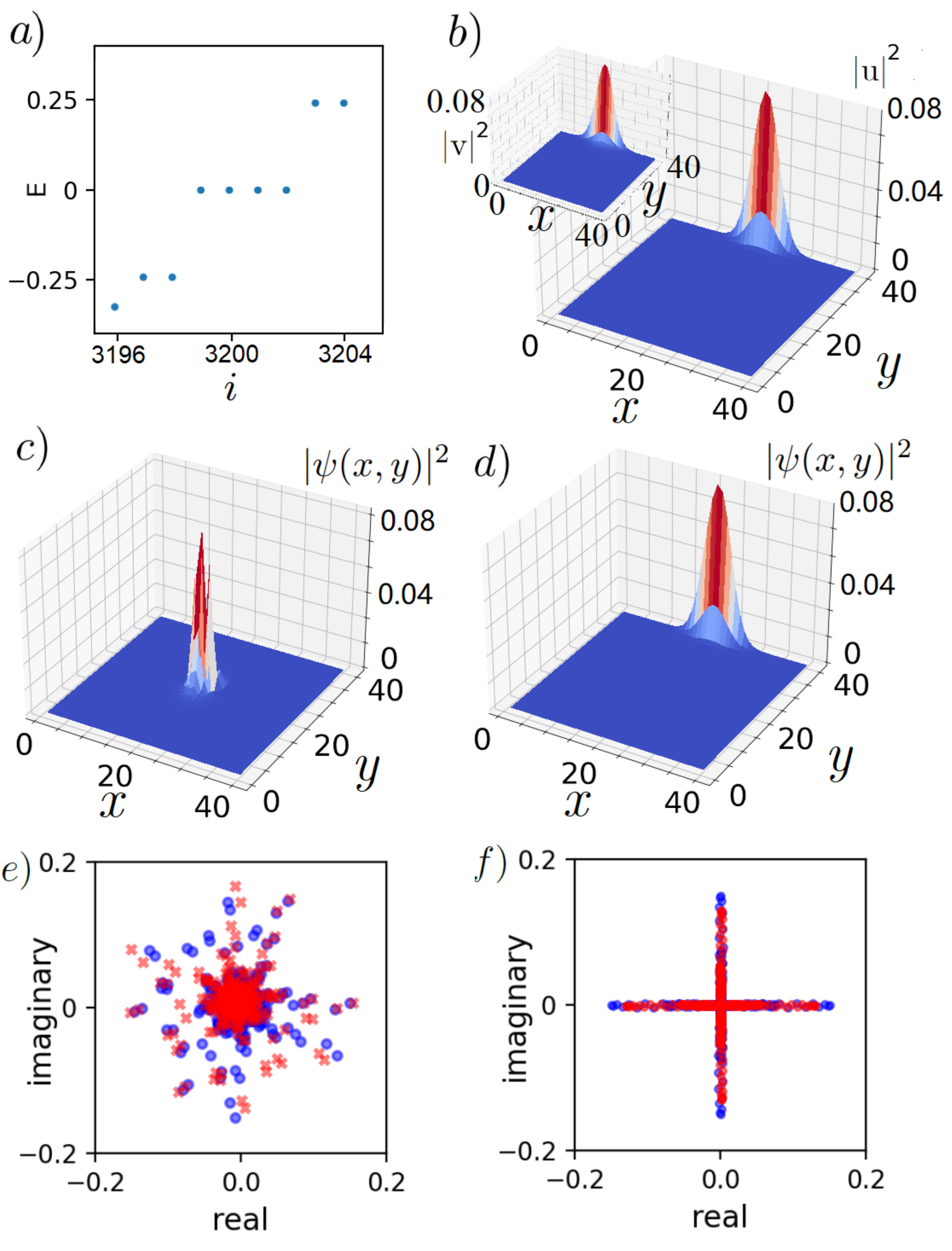}
\caption{For the disorder-averaged four-band Bloch Hamiltonian characterizing the 3D topological skyrmion phase and given by Eq.~\eqref{eq:3D}, constructed from the Hopf insulator Bloch Hamiltonian, its $\mc{C}'$ partner, and spin triplet pairing gap function given by Eq.~\eqref{eq:stp} with open boundary conditions in the $x$ and $y$ directions and the $z$ component of momentum $k_z$ replaced with $\theta$ characterizing a defect, (a) spectrum vs. eigenvalue index $i$ in the vicinity of zero-energy, showing a four-fold degeneracy at zero-energy. (b), (c) and (d) show the probability density for one zero-mode localized at the defect and one localized at the boundary, respectively. (e) and (f) show each complex number in the eigenvector plotted in the complex plane for a zero-energy state localized at the defect and at the boundary, respectively. }
\label{fig: Hopf4band}
\end{figure}

For this model, we then consider open boundary conditions in the $\hat{x}$- and $\hat{y}$-directions, and substitute an angle $\theta$ for momentum component $k_z$. The results for 3D chiral skyrmion phase Hamiltonian Eq.~\eqref{eq:Hopf 3D} under these conditions, when the 3D skyrmion number is $1$, are shown in Fig.~\ref{fig: Hopf4band} with additional disorder-averaging to illustrate that the results are independent of the specific form of $\Delta_t(\boldsymbol{k})$. The on-site disorder term takes the form of $\epsilon \sim \tau_3\sigma_0 \cdot \epsilon _0\textnormal{U}(-1,1)$ where $\tau_3\sigma_0$ is a $C'$ conserving term, $\epsilon _0$ is the magnitude of the disorder which is 5\% of the difference between non-zero energy levels (i.e. the lowest positive level and the highest negative level), and $\textnormal{U}(-1,1)$ is a uniform distribution with values from -1 to 1. The results shown are obtained from averaging 100 realizations.
\\

While two zero-energy states are realized for the Hopf insulator under these boundary conditions as shown in Fig.~\ref{fig: Hopf2band}(a), four zero-energy states are realized in the low-energy spectrum under the same conditions for the 3D topological skyrmion phase, as shown in Fig.~\ref{fig: Hopf4band}(a). The robustness of these states against local disorder in the latter case has been numerically verified by introducing random on-site perturbation at each lattice site.\\

Fig.~\ref{fig: Hopf4band}(b) shows the probability density profile of the particle component (denoted by $u$) and hole component (denoted by $v$) of a zero-energy eigenstate realized by the 3D topological skyrmion phase Hamiltonian, respectively. This can be compared directly with Fig.~\ref{fig: Hopf2band}(b). Notably, we see that, unlike in the case of the Hopf insulator, the zero-modes of the 3D topological skyrmion phase outputted by numerical diagonalization are particle-hole symmetric, but also localized.  Fig.~\ref{fig: Hopf4band}(c)(d) each show one localized zero-mode for direct comparison with Fig.~\ref{fig: Hopf2band}(c)(d). These two localized zero-mode wavefunctions at the defect and boundary are then visualized in the complex plane in Fig.~\ref{fig: Hopf4band}(e) and (f), respectively, similarly to Fig.~\ref{fig: Hopf2band}(e) and (f). That is, each wavefunction is represented by an array of $40 \times 40\times 4$ complex numbers, and these complex numbers are plotted in the complex plane as individual points. Similarly to the unpaired MZM at the defect in the Hopf insulator, the zero-modes localized at the defect in the 3D topological skyrmion phase have complex structure. The zero-modes of the 3D topological skyrmion phase localized at the boundary, however, reveal an interesting difference between the case of the Hopf insulator and the 3D topological skyrmion phase. They appear as crosses, or two straight, perpendicular lines, rather than as a single straight line as for a MZM, in the complex plane. In agreement with the low-energy theory presented in the following section and in the supplemental materials for the zero-modes on the boundary, linear combinations of these two-cross states yield two unpaired Majorana zero-modes (one purely real and one purely imaginary), each localized at the boundary. Rather than combine to form a generic complex fermion bound state, however, they are forbidden from hybridizing by a combination of $C'$ and $C$ symmetry to instead yield the cross structure.

\subsection{Analytical characterization of topologically-protected zero-modes}

To further characterize the cross-zero modes of the 3D topological skyrmion phase, we derive a low-energy effective theory for the model Hamiltonian Eq.~\eqref{eq:Hopf 3D} with open boundary conditions corresponding to an annulus geometry as in Yan~\emph{et al.}~\cite{yanzhongbo2017} and shown schematically in Fig.~\ref{fig:annulus}. Full details are presented in the Supplementary Materials, and here we focus on the resultant low-energy effective theory.

For a semi-infinite geometry, with the sample occupying the $x>0$ region the resultant effective Hamiltonian written in the on-site basis of the full Hamiltonian (generalized particle-hole DOF and spin DOF) is
\begin{align}
   H_{\textnormal{eff}} &=
   \begin{pmatrix}
    0 & \frac{3}{4}k_y+\frac{3}{4}i\lambda & -\frac{3}{8}\Delta_0 & 0\\
    \frac{3}{4}k_y-\frac{3}{4}i\lambda & 0 & 0 & \frac{3}{8}\Delta_0\\
    -\frac{3}{8}\Delta_0 & 0 & 0 & -\frac{3}{4}k_y + \frac{3}{4}i\lambda\\
    0 & \frac{3}{8}\Delta_0 & -\frac{3}{4}k_y - \frac{3}{4}i\lambda & 0
    \end{pmatrix}
\end{align}
By taking $k_y\rightarrow -i\partial_y$ and $\lambda \rightarrow y/R$, $H_{\textnormal{eff}}$ becomes
\begin{align}
   H_{\textnormal{eff}} &= \frac{3}{4}
   \begin{pmatrix}
    0 & -i\partial_y+i\frac{1}{R}y & -\frac{1}{2}\Delta_0 & 0\\
    -i\partial_y-i\frac{1}{R}y & 0 & 0 & \frac{1}{2}\Delta_0\\
    -\frac{1}{2}\Delta_0 & 0 & 0 & i\partial_y+i\frac{1}{R}y\\
    0 & \frac{1}{2}\Delta_0 & i\partial_y-i\frac{1}{R}y & 0
    \end{pmatrix}.
\end{align}

Squaring both sides of the time-independent Schrödinger equation yields the resultant equation
\begin{align}\frac{9}{16}&
    \begin{pmatrix}
    \epsilon_{y,-} & 0 & 0 & -i\partial_y \Delta_0\\
    0 & \epsilon_{y,+} & i\partial_y \Delta_0 & 0\\
    0 & i\partial_y \Delta_0 & \epsilon_{y,+} & 0\\
    -i\partial_y \Delta_0 & 0 & 0 & \epsilon_{y,-}
    \end{pmatrix} \psi = E^2 \psi,
\end{align}
where $\epsilon_{y,\pm} = -\partial_y^2+\frac{1}{R^2}y^2\pm \frac{1}{R}+\frac{1}{4}\Delta_0^2$.

The eigenstates of this low-energy effective Hamiltonian may then be expressed as $|\mu_i \rangle = \chi_i^{\top} |\psi_i \rangle$, where $\{\chi_i \}$ ($i \in \{1,2,3,4\}$) are eigenvectors of $\tau_2\sigma_2$, and $|\psi_i\rangle = |\psi_i(y)\rangle$ is a spatially-varying scalar function satisfying the $i$\textsuperscript{th} differential equation of the set of four differential equations contained in the Hamiltonian. This yields $|\mu_i\rangle$ which take the forms
\begin{equation}
\begin{aligned}
\label{eigenstates}
    |\mu_1\rangle &= \frac{1}{\sqrt{2}}(1,0,0,1)^\top |\psi_1\rangle&
    |\mu_2\rangle &= \frac{1}{\sqrt{2}}(0,-1,1,0)^\top |\psi_2\rangle\\
    |\mu_3\rangle &= \frac{1}{\sqrt{2}}(-1,0,0,1)^\top |\psi_3\rangle&
    |\mu_4\rangle &= \frac{1}{\sqrt{2}}(0,1,1,0)^\top |\psi_4\rangle
\end{aligned}
\end{equation}

These states serve as zero-energy eigenstates of the effective low-energy Hamiltonian if the following differential equations are satisfied by the $\{\psi_i(y) \}$ (omitting $\frac{1}{4}\Delta_0^2$ as higher order terms):
\begin{align}
    \label{DE1}
    \left(-\partial_y^2+\frac{1}{R^2}y^2-\frac{1}{R}-i\partial_y \Delta_0\right)|\psi_1\rangle &= 0\\
    \label{DE2}
    \left(-\partial_y^2+\frac{1}{R^2}y^2+\frac{1}{R}-i\partial_y \Delta_0\right)|\psi_2\rangle &= 0\\
    \label{DE3}
    \left(-\partial_y^2+\frac{1}{R^2}y^2-\frac{1}{R}+i\partial_y \Delta_0\right)|\psi_3\rangle &= 0\\
    \label{DE4}
    \left(-\partial_y^2+\frac{1}{R^2}y^2+\frac{1}{R}+i\partial_y \Delta_0\right)|\psi_4\rangle &= 0.
\end{align}
Note that by setting $\Delta_0=0$ we recover the differential equation in the two-band Hopf insulator which only has a solution for $-\frac{1}{R}$, so that we can take this solution, treat $-i\partial _y \Delta_0$ as a perturbation term, apply first order perturbation theory (for degenerate energy levels), and obtain the spatial distribution of the zero-energy modes, $|\psi_1 (y) \rangle$.

The zero-energy eigenstates at the inner boundary then take the form,
\begin{align}
    |0\rangle_R   &=|\psi_{\org{3}}(y)\rangle
                \begin{pmatrix}
                1  \\
                0 \\
                0\\
                -1\\
                \end{pmatrix}\\
    |0\rangle_I   &=|\psi_{\org{1}}(y)\rangle
                \begin{pmatrix}
                i  \\
                0 \\
                0\\
                i\\
                \end{pmatrix}.
\end{align}

The above analysis is carried out for $x>0$ i.e. the inner boundary. When instead considering the other boundary corresponding to $x<0$, we reach the same differential equations \eqref{DE1} and \eqref{DE3}, this time with eigenstates $|\mu_2\rangle$ and $|\mu_4\rangle$ in Eqn~\eqref{eigenstates}. Hence, we similarly find two additional zero-energy states of the form,
\begin{align}
    |\bar{0}\rangle_R   &=|\psi_{\org{1}}(y)\rangle
                \begin{pmatrix}
                1  \\
                0 \\
                0\\
                -1\\
                \end{pmatrix}\\
    |\bar{0}\rangle_I   &=|\psi_{\org{3}}(y)\rangle
                \begin{pmatrix}
                i  \\
                0 \\
                0 \\
                i \\
                \end{pmatrix},
\end{align}
where $|\psi_2(y) \rangle$ is a corresponding spatial distribution for states at the outer boundary.

These results indicate that $\mathcal{C}'$ symmetry alone yields two Majorana zero-modes at the defect and two at the boundary, but this is insufficient to realize the cross zero-modes. $\mc{C}'$ symmetry in combination with $\mc{C}$ symmetry of the Hopf insulator Bloch Hamiltonian is also required: the cross states are the most general complex states permitted by $\mc{C}$ and $\mc{C}'$ symmetry. Specifically, certain linear combinations of these four cross states satisfy constraints imposed by $\mc{C}$, that the four zero-modes may be considered as two pairs of MZMs, with one MZM in each pair localized at the defect and the other localized at the boundary, but may also instead be considered as two other pairings of MZMs as required by $\mc{C}'$ symmetry, with one pair localized at the boundary and one pair localized at the defect. This difference stems from the relationship between $\mc{C}$ and $\mc{C}'$, that $\mc{C}'$ acts like $\mc{C}$ upon exchange of position and momentum variables. This point was, to our knowledge, first observed in work by Liu~\emph{et al.}~\cite{liu2017}: under $\mc{C}$ transformation, the position variables are invariant, but all of the momenta variables pick up a sign. However, under the $\mc{C}'$ transformation, the momenta do not change, but all of the position variables pick up a sign. Therefore, for all intents and purposes, we can replace $\mc{C}'$ with $\mc{C}$ as long as we reverse the roles of the position and momenta variables.

We may also understand the non-trivial topological state from the perspective of the homotopy group $\pi_3\left(\mathrm{Sp}(2N)  / \mathrm{U}(N)\right) = \mathbb{Z}_2$ being relevant both to mappings from the Brillouin zone to the space of flat-band Hamiltonians, but also to mappings from the Brillouin zone to the space of ground state spin expectation value yielding the topological momentum-space, ground-state spin textures discussed here. This could correspond to a $\mathbb{Z}_2 \times \mathbb{Z}_2$ topological classification for the system overall. As a result, even if the topological invariant associated with mappings from the Brillouin zone to the space of flat-band Hamiltonians is even and trivial, the system may still harbor topologically-protected zero-modes due to the co-existing topological skyrmion phase.

This Hamiltonian realizing the cross zero-modes may therefore be interpreted as a \textit{generalized} BdG Hamiltonian for a centrosymmetric superconductor, where $\mc{C}$ and $\mc{C}'$ \textit{do not transform} the normal state Bloch Hamiltonian in the same way because it is not even in momentum. For such cases, the Hilbert space is enlarged with both a particle-hole DOF, and an additional \textit{generalized} particle-hole DOF for $\mc{C}'$. We can therefore also interpret the full Hamiltonian as a spinless superconductor, written in terms of the Bloch Hamiltonian as $\mc{H} = \sum_{\bk} \Psi^{\dagger}_{\bk} \mc{H}(\bk) \Psi_{\bk}/2$,
where $\Psi_{\bk} = \left(c^{}_{\bk}, c^{\dagger}_{-\bk}, c^{\dagger}_{\bk},  c^{}_{-\bk} \right)^t$. Here, $c^{}_{\bk}$ is the annihilation operator of electrons with momentum $\bk = \left( k_x, k_y, k_z \right)$.

It is interesting that the cross zero-modes are strongly localized only at one boundary or the other: this is due to only two linearly-independent cross states being possible at a given point in real-space. The cross-states therefore must additionally be orthogonal as a result of their spatial distribution. Therefore two cross states are strongly localized at the inner boundary, and the remaining two are strongly localized at the outer boundary to form a four-fold degenerate ground state manifold. Importantly, this indicates the cross zero-modes should be more robust against finite-size effects than unpaired Majorana zero-modes. This is supported by numerics, which show well-defined cross zero-modes at the outer boundary in smaller systems (see \ref{fig: Hopf4band} (f)).

In summary, in the topologically non-trivial regime of the Hamiltonian presented here realizing the 3D topological skyrmion phase, corresponding to a skyrmion forming in the ground-state spin expectation value texture at half-filling, there are two cross states localized on the outer boundary, and two other cross states localized at the defect. These cross states are a manifestation of a bulk-boundary correspondence of topological skyrmion phases due to topology of defects trapped by the momentum-space spin textures.

\section{Discussion \& Conclusion}\label{4}
In this work, we present toy models for recently-identified topological phases of matter in two dimensions, characterized by the topological charge of a momentum-space spin texture, a \textit{spin topological invariant} rather than the topological charge of a momentum-space texture of the projector onto occupied states, a \textit{projector topological invariant}, as done in the ten-fold way classification scheme~\cite{schnyder_2008}. We realize these topological phases in systems which can therefore be more fully characterized in terms of both a projector topological invariant and a spin topological invariant. Extending our results to introduce three-dimensional topological skyrmion phases corresponding to trivial projector topological invariant and non-trivial spin topological invariant, we furthermore find evidence of a bulk-boundary correspondence of topological skyrmion phases. As bulk-boundary correspondence is generally expected only for non-trivial projector topological invariant, these results serve as an important signature and consequence of topological skyrmion phases.

The bulk-boundary correspondence considered here, a \textit{defect bulk-boundary correspondence}, is realized by opening boundary conditions of the 3D system in two directions $x$ and $y$, and replacing the momentum component in a third $z$ direction by an angle $\theta(x,y)$, which can characterize a zero-dimensional defect in the $xy$-plane.  For these open boundary conditions, we characterize the system's topology in terms of a bulk projector invariant, the Hopf invariant, and a bulk spin invariant, the 3D skyrmion number. Under these open boundary conditions, the topological skyrmion phases corresponding to trivial projector invariant and non-trivial spin invariant possess topologically-protected ``cross'' zero-modes, which generalize unpaired Majorana zero-modes: the zero-mode wavefunctions in lattice-based systems consist of a set of complex numbers that form a cross-like structure when each of these complex numbers is plotted in the complex plane, rather than forming a straight line corresponding to a purely real state up to some phase degree of freedom as in the case of an unpaired Majorana zero-mode.

In order to demonstrate this defect bulk-boundary correspondence of the topological skyrmion phases, we first present a method for constructing Bloch Hamiltonian models for topological skyrmion phases with four bands and characterize myriad examples. We furthermore show BdG Hamiltonians for centrosymmetric superconductors naturally possess the $\mc{C}'$ symmetry in addition to particle-hole symmetry $\mc{C}$ when the normal state Hamiltonian is even in momentum. More generally, the topological skyrmion phases are described by generalized BdG Hamiltonians corresponding to artificially enlarging the Hilbert space to possess a generalized particle-hole DOF in addition to a particle-hole DOF.

We introduce an efficient method for computing skyrmion numbers based on past work by Fukui~\emph{et al.}~\cite{fukui2005}. At the four-band level, we find the skyrmion number computed using the ground state spin expectation value at half-filling is no longer equal to the total Chern number. Instead, we find that the total Chern number $\mc{C}$ and the skyrmion number $\mc{Q}$ are related by $\mc{C} = -2\mc{Q}$ analytically for the atomic model---and numerically for all other models considered---of the chiral topological skyrmion phase characterized by Bloch Hamiltonians with four bands, when each of $\mc{C}$ and $\mc{Q}$ are topologically stable. It is also possible for $\mc{C}$ to be topologically stable and $\mc{Q}$ to be topologically unstable, due to the magnitude of the ground state spin being zero somewhere in the Brillouin zone. We do not find any instances where the reverse case occurs, in which $\mc{Q}$ is topologically stable and $\mc{C}$ is topologically unstable, and will explore this further in future work.

We do not explore bulk-boundary correspondence of two-dimensional topological skyrmion phases in this work. However, since completion of this work, bulk-boundary correspondence for the 2D topological skyrmion phases has also been further characterized in terms of the observable-enriched partial trace and observable-enriched entanglement spectrum introduced by Winter~\emph{et al.}~\cite{Cook2022, Winter2023}. That is, bulk-boundary correspondence associated with non-trivial skyrmion number $\mc{Q}$, when boundary conditions are opened in one direction, corresponds to $\mc{Q}$ chiral modes in the observable-enriched entanglement spectrum bulk gap.

 We use knowledge from constructing various toy models for the chiral topological skyrmion phase in two dimensions to introduce the three-dimensional topological skyrmion phase, constructing it from the Hopf insulator Hamiltonian. We characterize the topology of these models using both a projector topological invariant and a spin topological invariant. We then show that the three-dimensional spin skyrmion in momentum space traps a defect, which can yield a bulk-boundary correspondence for the three-dimensional skyrmion phase: taking a momentum component to be an angle $\theta$ characterizing a real-space defect in a two-dimensional system with open boundary conditions in the other two directions, we find two zero-modes are localized at the defect and two are localized at the boundary when the system is in the three-dimensional topological skyrmion phase. These zero-modes are realized when the projector topological invariant, the Hopf invariant, is trivial and the spin topological invariant, the 3D skyrmion number, is non-trivial. Therefore, while bulk-boundary correspondence has previously been associated strictly with non-trivial projector topological invariants, we show bulk-boundary correspondence is also associated with non-trival spin topological invariant. Such a zero-mode exhibits a cross-like structure when each of the complex numbers characterizing the zero-mode wavefunction is plotted in the complex plane. Such a cross zero-mode is protected by a combination of particle-hole and generalized particle-hole symmetries and is robust against disorder. They generalize unpaired Majorana zero-modes, and may be thought of as an unpaired Majorana zero-mode and its $\mc{C}'$ partner combined in a single, localized state, yet isolated from one another by a $\pi/2$ phase difference in the complex plane.

In conclusion, these results indicate that the topological skyrmion phases exhibit a bulk-boundary correspondence by trapping defects that can yield topologically-protected, gapless boundary states. This serves as confirmation of the topological skyrmion phases and their implications, including their type-II topological phase transition contradicting the flat-band limit assumption. As the flat-band limit assumption is used in the construction of the ten-fold way topological classification scheme of Schnyder et al.~\cite{schnyder_2008} and Ryu et al.~\cite{Ryu_2010}, the type-II topological phase transition of the topological skyrmion phases~\cite{Cook2022} indicates topological skyrmion phases are a new topological class outside the classification of
Schnyder et al.~\cite{schnyder_2008} and Ryu et al.~\cite{Ryu_2010}, rather than
additional topological characterization of the current topological
classification of free fermions. In addition, these gapless boundary states due to defect bulk-boundary correspondence of the topological skyrmion phases may interact with the boundary states of other topological phases. Such interplay between myriad topological phases will be explored in future work, especially given the potential for the topological skyrmion phase and the Chern insulator phase to co-exist. Finally, the properties and consequences of the cross zero-modes of the three-dimensional topological skyrmion phase will also be explored further in future work, given the significance of unpaired Majorana zero-modes for topological quantum computation and the fundamental generalization of unpaired Majorana zero-modes presented by the cross zero-modes discovered here.

\textbf{Acknowledgements} -  The authors thank B. Braunecker and B. Doucot for helpful discussions and comments on the manuscript.

\bibliographystyle{apsrev4-2}
\bibliography{p1bib.bib}

%\appendix*
%\input{sections/appendix1.tex}

\clearpage

%%%%%%%%% Prefix a "S" to all equations, figures, tables and reset the counter %%%%%%%%%%
\makeatletter
\renewcommand{\theequation}{S\arabic{equation}}
\renewcommand{\thefigure}{S\arabic{figure}}
\renewcommand{\thesection}{S\arabic{section}}
\setcounter{equation}{0}
\setcounter{section}{0}
\onecolumngrid
\begin{center}
  \textbf{\large Supplemental material for ``Defect bulk-boundary correspondence of topological skyrmion phases of matter''}\\[.2cm]
  Shu-Wei Liu,$^{1}$ Li-kun Shi,$^{1}$ and Ashley M. Cook$^{1,2,*}$\\[.1cm]
  {\itshape ${}^1$Max Planck Institute for Chemical Physics of Solids, Nöthnitzer Strasse 40, 01187 Dresden, Germany\\
  ${}^2$Max Planck Institute for the Physics of Complex Systems, Nöthnitzer Strasse 38, 01187 Dresden, Germany\\}
  ${}^*$Electronic address: cooka@pks.mpg.de\\
(Dated: \today)\\[1cm]
\end{center}

\section{Low-energy, analytical theory of the 3D topological skyrmion phase}

Here we derive a low-energy theory for the four-band, three-dimensional skyrmion phase with Hamiltonian defined in Eq.~(34), for a semi-infinite geometry with the sample occupying the $x>0$ region. Translational symmetry in the $\hat{x}$-direction is therefore broken while momentum component $k_y$ remains as a good quantum number. Thus, a real-space coordinate is used in the $\hat{x}$-direction to label lattice sites. The wave functions and energy eigenvalues can be obtained by solving the following equation
\begin{equation}
    \begin{pmatrix}
    H_0 & H_1 & H_2 & 0 & 0 & 0 & \cdots \\
    H_1^\dagger & H_0 & H_1 & H_2 & 0 & 0 & \cdots \\
    H_2^\dagger & H_1^\dagger & H_0 & H_1 & H_2 & 0 & \cdots \\
    0 & H_2^\dagger & H_1^\dagger & H_0 & H_1 & H_2 & \cdots \\
    \vdots & \vdots & \vdots & \vdots & \vdots & \vdots & \ddots \\
    \end{pmatrix}
    \begin{pmatrix}
    \psi_1 \\
    \psi_2 \\
    \psi_3 \\
    \psi_4 \\
    \vdots
    \end{pmatrix}
    =
    E
    \begin{pmatrix}
    \psi_1 \\
    \psi_2 \\
    \psi_3 \\
    \psi_4 \\
    \vdots
    \end{pmatrix}
    \label{eqn:main}
\end{equation}
where $\psi_j$ is the $j^{th}$ entry of an eigenstate for this equation corresponding to the $j$\textsuperscript{th} lattice site in the $\hat{x}$-direction. Each $\psi_j$ has four components corresponding to the generalized particle-hole and spin degrees of freedom at each site in real-space, and the non-zero terms of the Hamiltonian matrix representation on the left-hand side of Eq.~\ref{eqn:main} take the following forms: the term on the diagonal is

\begin{equation}
    H_0(k_y,\lambda,\Delta_0)
    =
    \begin{pmatrix}
    h_0 & 0\\
    0 & -h_0^*\\
    \end{pmatrix}
    + \Delta_s,
\end{equation}
and terms off the diagonal are

\begin{equation}
     H_1(k_y,\lambda,\Delta_0)
    =
    \begin{pmatrix}
    h_1 & 0\\
    0 & -h_1^*\\
    \end{pmatrix}
    +
    \begin{pmatrix}
    0 & 0 & -\frac{1}{2}\Delta_0 & 0\\
    0 & 0 & 0 & -\frac{1}{2}\Delta_0\\
    \frac{1}{2}\Delta_0 & 0 & 0 & 0\\
    0 & \frac{1}{2}\Delta_0 & 0 & 0
    \end{pmatrix},
\end{equation}
and
\begin{equation}
    \begin{aligned}
    H_2
    &=
    \begin{pmatrix}
    -\frac{1}{2} & -\frac{1}{2} & 0 & 0\\
    \frac{1}{2} & \frac{1}{2} & 0 & 0\\
    0 & 0 & \frac{1}{2} & \frac{1}{2}\\
    0 & 0 & -\frac{1}{2} & -\frac{1}{2}
    \end{pmatrix},
    \end{aligned}
\end{equation}
where $\lambda$ is a parameter substituted for the $\hat{z}$-component of the momentum, $k_z$, characterizing a defect. Here, $\lambda=n\theta(i,j)$ with $\theta$ being the relative angle between the defect and site$(i,j)$, $n$ is an integer taken to be $1$ in this study and $\Delta_0$ is the strength of the spin-triplet pairing.

Here, $h_0$ and $h_1$ are momentum- and $\lambda$-dependent terms taking the forms
\begin{equation}
\begin{aligned}
    h_0(k_y,\lambda)
    &=
    \begin{pmatrix}
    A(k_y,\lambda) & B(k_y,\lambda)\\
    B(k_y,\lambda) & -A(k_y,\lambda)
    \end{pmatrix}\\
    h_1(k_y,\lambda)
    &=\begin{pmatrix}
    \frac{3}{2}-\cos{k_y}-\cos{\lambda}
    & -i\sin{\lambda}+\sin{k_y}-(\cos{k_y}+\cos{\lambda}-\frac{3}{2})\\
    -i\sin{\lambda}+\sin{k_y}+(\cos{k_y}+\cos{\lambda}-\frac{3}{2})
    & -\frac{3}{2}+\cos{k_y}+\cos{\lambda}
    \end{pmatrix}\\
\end{aligned}
\end{equation}
where here
\begin{equation}
\begin{aligned}
    A(k_y,\lambda)&=-\frac{13}{4}-\cos{2k_y}+3(\cos{k_y}+\cos{\lambda})-2\cos{k_y}\cos{\lambda}\\
    B(k_y,\lambda)&=2i\sin{k_y}\sin{\lambda}+(\sin{2k_y}+2\sin{k_y}\cos{\lambda}-3\sin{k_y}).
\end{aligned}
\end{equation}

$\Delta_s$ is also momentum-dependent, taking the form
\begin{equation}
\begin{aligned}
    \Delta_s(k_y,\Delta_0)
    &=
    \begin{pmatrix}
    0 & 0 & -\Delta_0 \sin{k_y} & 0\\
    0 & 0 & 0 & \Delta_0 \sin{k_y}\\
    -\Delta_0 \sin{k_y} & 0 & 0 & 0\\
    0 & \Delta_0 \sin{k_y} & 0 & 0
    \end{pmatrix}.
\end{aligned}
\end{equation}

To find zero-energy solutions in analogy to Yan~\emph{et al.}~\cite{yanzhongbo2017}, we first set $k_y=0$, $\lambda=0$ and $\Delta_0=0$. The non-zero entries of the Hamiltonian in Eq.~\ref{eqn:main} then take the following forms:
\begin{equation}
\begin{aligned}
    H_0(0,0,0)
    &=
    \begin{pmatrix}
    -\frac{1}{4} & 0 & 0 & 0\\
    0 & \frac{1}{4} & 0 & 0\\
    0 & 0 & \frac{1}{4} & 0\\
    0 & 0 & 0 & -\frac{1}{4}
    \end{pmatrix}\\
    H_1(0,0,0)
    &=
    \begin{pmatrix}
    -\frac{1}{2} & -\frac{1}{2} & 0 & 0\\
    \frac{1}{2} & \frac{1}{2} & 0 & 0\\
    0 & 0 & \frac{1}{2} & \frac{1}{2}\\
    0 & 0 & -\frac{1}{2} & -\frac{1}{2}
    \end{pmatrix}
    =
    H_2
\end{aligned}
\end{equation}

We then consider eigenstates of

\begin{equation}
    \tau_0 \otimes \sigma_1
    =
    \begin{pmatrix}
    0 & 1 & 0 & 0\\
    1 & 0 & 0 & 0\\
    0 & 0 & 0 & 1\\
    0 & 0 & 1 & 0
    \end{pmatrix},
\end{equation}
which are
\begin{equation}
\begin{aligned}
    |\nu_1\rangle &= \frac{1}{\sqrt{2}}(0,0,-1,1)^\top \\
    |\nu_2\rangle &= \frac{1}{\sqrt{2}}(-1,1,0,0)^\top \\
    |\nu_3\rangle &= \frac{1}{\sqrt{2}}(0,0,1,1)^\top \\
    |\nu_4\rangle &= \frac{1}{\sqrt{2}}(1,1,0,0)^\top
\end{aligned}
\end{equation}
 where $|\nu_1\rangle$ and $|\nu_2\rangle$ have eigenvalue $\lambda_{-}=-1$, and $|\nu_3\rangle$ and $|\nu_4\rangle$ have eigenvalue $\lambda_{+}=+1$. Operating the matrices $H_i$'s on the $|\nu_i\rangle$'s, they produce:
\begin{equation}
\begin{aligned}
    H_0
    \begin{pmatrix}
    |\nu_1\rangle \\
    |\nu_2\rangle \\
    |\nu_3\rangle \\
    |\nu_4\rangle \\
    \end{pmatrix}
    &=
    \frac{1}{4}
    \begin{pmatrix}
    -|\nu_3\rangle \\
    |\nu_4\rangle \\
    -|\nu_1\rangle \\
    |\nu_2\rangle \\
    \end{pmatrix}\\
    H_1
    \begin{pmatrix}
    |\nu_1\rangle \\
    |\nu_2\rangle \\
    |\nu_3\rangle \\
    |\nu_4\rangle \\
    \end{pmatrix}
    &=
    \begin{pmatrix}
    0 \\
    0 \\
    -|\nu_1\rangle \\
    |\nu_2\rangle \\
    \end{pmatrix}\\
    H_1^\dagger
    \begin{pmatrix}
    |\nu_1\rangle \\
    |\nu_2\rangle \\
    |\nu_3\rangle \\
    |\nu_4\rangle \\
    \end{pmatrix}
    &=
    \begin{pmatrix}
    -|\nu_3\rangle \\
    |\nu_4\rangle \\
    0 \\
    0 \\
    \end{pmatrix}
\end{aligned}
\end{equation}
The zero-energy wavefunctions take the form
\begin{equation}
    |\Psi_i\rangle = \sum_j a_{i,j} |j\rangle \otimes |\nu_i\rangle
\end{equation}
where $a_{i,j}$ is a function of $j$ that normalizes $|\Psi_i\rangle$ and $|j\rangle$ indicates localization on the $j^{th}$ site:
\begin{equation}
    |j\rangle = (0,0,\cdots,0,1,0,\cdots)^\top
\end{equation}
i.e. having $1$ in the $j^{th}$ element and $0$ everywhere else. We can explore different possibilities of $|\Psi_i\rangle$ by substituting this expression in Eq.\eqref{eqn:main} and consider the $j^{th}$ element of the equation. In the case of $i=1$:
\begin{equation}
    H_2^\dagger a_{1,j-2} |\nu_1\rangle + H_1^\dagger a_{1,j-1} |\nu_1\rangle + H_0 a_{1,j} |\nu_1\rangle + H_1 a_{1,j+1} |\nu_1\rangle + H_2 a_{1,j+2} |\nu_1\rangle = E = 0
\end{equation}
which simplifies to
\begin{equation}
    (-a_{1,j-2}-a_{1,j-1}-\frac{1}{4}a_{1,j}+0+0)|\nu_3\rangle = 0
\end{equation}
and therefore obtaining the following recurrence relation in $a_{1,j}$:
\begin{equation}
    a_{1,j} = -4(a_{1,j-2}+a_{1,j-1})
\end{equation}
which solves to give the general formula:
\begin{equation}
    a_{1,j} = (-2)^j C_1 + (-2)^j j C_2
\end{equation}
but there is no normalizable solution for $C_1$ or $C_2$. \\
In the case of $i=2$:
\begin{equation}
    H_2^\dagger a_{2,j-2} |\nu_2\rangle + H_1^\dagger a_{2,j-1} |\nu_2\rangle + H_0 a_{2,j} |\nu_2\rangle + H_1 a_{2,j+1} |\nu_2\rangle + H_2 a_{2,j+2} |\nu_2\rangle = E = 0
\end{equation}
which simplifies to
\begin{equation}
    (a_{2,j-2}+a_{2,j-1}+\frac{1}{4}a_{1,j}+0+0)|\nu_4\rangle = 0
\end{equation}
and therefore obtaining the following recurrence relation in $a_{2,j}$:
\begin{equation}
    a_{2,j} = -4(a_{2,j-2}+a_{2,j-1}).
\end{equation}
which is identical to the $i=1$ case and is unnormalizable.

In the case of $i=3$:
\begin{equation}
    H_2^\dagger a_{3,j-2} |\nu_3\rangle + H_1^\dagger a_{3,j-1} |\nu_3\rangle + H_0 a_{3,j} |\nu_3\rangle + H_1 a_{3,j+1} |\nu_3\rangle + H_2 a_{3,j+2} |\nu_3\rangle = E = 0
\end{equation}
which simplifies to
\begin{equation}
    (0+0-\frac{1}{4}a_{3,j}-a_{2,j+1}-a_{2,j+2})|\nu_1\rangle = 0
\end{equation}
and therefore obtaining the following recurrence relation in $a_{3,j}$:
\begin{equation}
    a_{3,j} = -4(a_{3,j+1}+a_{3,j+2}).
\end{equation}
which solves to give the general formula:
\begin{equation}
    a_{3,j} = (-\frac{1}{2})^j C_1 + (-\frac{1}{2})^j j C_2
\end{equation}
which now has normalizable solution for $C_1$ and $C_2$. Note that each of the two terms in $a_{3,j}$ can form an individual solution, so we denote the two choices of $a_{3,j}$ as $\alpha_j$ and $\beta_j$, and thus obtain the following expression:
\begin{equation}
\begin{aligned}
    \alpha_j &= \sqrt{3}(-\frac{1}{2})^j\\
    \beta_j &= -\sqrt{\frac{27}{20}}(-\frac{1}{2})^{j-1}(j-1)
\end{aligned}
\end{equation}
and in summary:
\begin{equation}
\begin{aligned}
        |\Psi_\alpha\rangle &= \sum_j \alpha_{j} |j\rangle \otimes |\nu_3\rangle\\
        |\Psi_\beta\rangle &= \sum_j \beta_{j} |j\rangle \otimes |\nu_3\rangle\\
\end{aligned}
\end{equation}
such that
$\langle\Psi_\alpha|\Psi_\alpha\rangle = \langle\Psi_\beta|\Psi_\beta\rangle=1.$

In the case of $i=4$:
\begin{equation}
    H_2^\dagger a_{4,j-2} |\nu_4\rangle + H_1^\dagger a_{4,j-1} |\nu_4\rangle + H_0 a_{4,j} |\nu_2\rangle + H_1 a_{4,j+1} |\nu_4\rangle + H_2 a_{4,j+2} |\nu_4\rangle = E = 0
\end{equation}
which simplifies to
\begin{equation}
    (0+0+\frac{1}{4}a_{4,j}+a_{2,j+1}+a_{2,j+2})|\nu_2\rangle = 0
\end{equation}
and therefore obtaining the following recurrence relation in $a_{4,j}$:
\begin{equation}
    a_{4,j} = -4(a_{4,j+1}+a_{4,j+2}).
\end{equation}
which is an identical form to the $i=3$ case so the solutions are
\begin{equation}
\begin{aligned}
    \gamma_j &= \sqrt{3}(-\frac{1}{2})^j\\
    \delta_j &= -\sqrt{\frac{27}{20}}(-\frac{1}{2})^{j-1}(j-1)
\end{aligned}
\end{equation}
and
\begin{equation}
\begin{aligned}
        |\Psi_\gamma\rangle &= \sum_j \gamma_{j} |j\rangle \otimes |\nu_4\rangle\\
        |\Psi_\delta\rangle &= \sum_j \delta_{j} |j\rangle \otimes |\nu_4\rangle\\
\end{aligned}
\end{equation}
such that
$\langle\Psi_\gamma|\Psi_\gamma\rangle = \langle\Psi_\delta|\Psi_\delta\rangle=1.$
The next step in normalization is to notice $|\Psi_\alpha\rangle$ and $|\Psi_\beta\rangle$ are both constructed from $|\nu_3\rangle$ so $\langle\Psi_\alpha|\Psi_\beta\rangle$ is not automatically 0. We apply the Gram-Schmidt orthogonalization as the following:
\begin{equation}
\begin{aligned}
    |\Psi_\alpha^o\rangle &= |\Psi_\alpha\rangle \\
    |\Psi_\beta^o\rangle &= N(|\Psi_\beta\rangle - |\Psi_\alpha\rangle \langle\Psi_\alpha|\Psi_\beta\rangle)
\end{aligned}
\end{equation}
so that $\langle\Psi_\alpha^o|\Psi_\beta^o\rangle$ is conveniently 0 and the normalization constant $N$ can be worked out from first determining
\begin{equation}
\begin{aligned}
    \langle\Psi_\alpha|\Psi_\beta\rangle &= \sum_j \alpha^*_j \beta_j\\
    &= \sum_j -\sqrt{\frac{81}{20}}(-\frac{1}{2})^{2j-1}(j-1)\\
    &=-\sqrt{\frac{81}{20}}(-\frac{2}{9})\\
    &= \frac{1}{\sqrt{5}}
\end{aligned}
\end{equation}
and then by requiring
\begin{equation}
\begin{aligned}
    \langle\Psi_\beta^o|\Psi_\beta^o\rangle &= N^2 (\langle\Psi_\beta|-\frac{1}{\sqrt{5}}\langle\Psi_\alpha|)(|\Psi_\beta\rangle-\frac{1}{\sqrt{5}}|\Psi_\alpha\rangle) = 1\\
    N^2(1-\frac{1}{5}-\frac{1}{5}+\frac{1}{5}) &= 1\\
    N &= \frac{\sqrt{5}}{2}
\end{aligned}
\end{equation}
and therefore
\begin{equation}
\begin{aligned}
    |\Psi_\alpha^o\rangle &= |\Psi_\alpha\rangle \\
    |\Psi_\beta^o\rangle &= \frac{\sqrt{5}}{2}|\Psi_\beta\rangle - \frac{1}{2}|\Psi_\alpha\rangle
\end{aligned}
\end{equation}
and similarly
\begin{equation}
\begin{aligned}
    |\Psi_\gamma^o\rangle &= |\Psi_\gamma\rangle \\
    |\Psi_\delta^o\rangle &= \frac{\sqrt{5}}{2}|\Psi_\delta\rangle - \frac{1}{2}|\Psi_\gamma\rangle.
\end{aligned}
\end{equation}

Next, we consider the Hamiltonian in the neighbourhood of $k_y=0, \lambda=0, \Delta_0=0$ by expanding $H$ to first order of these three variables i.e. $H_i(k_y,\lambda,\Delta_0)=H_i(0,0,0)+\Delta H_i(k_y,\lambda,\Delta_0)$ where
\begin{equation}
    \Delta H = \begin{pmatrix}
    \Delta H_0 & \Delta H_1 & 0 & 0 & 0 & \cdots \\
    \Delta H_1^\dagger & \Delta H_0 & \Delta H_1 & 0 & 0 & \cdots \\
    0 & \Delta H_1^\dagger & \Delta H_0 & \Delta H_1 & 0 & \cdots \\
    \vdots & \vdots & \vdots & \vdots & \vdots & \ddots \\
    \end{pmatrix}
\end{equation}
with
\begin{align}
    \Delta H_0
    &=
    \begin{pmatrix}
    0 & k_y & 0 & 0\\
    k_y & 0 & 0 & 0\\
    0 & 0 & 0 & -k_y\\
    0 & 0 & -k_y & 0
    \end{pmatrix}\\
    \Delta H_1
    &=
    \begin{pmatrix}
    0 & -i\lambda+k_y & -\frac{1}{2}\Delta_0 & 0\\
    -i\lambda+k_y & 0 & 0 & -\frac{1}{2}\Delta_0\\
    \frac{1}{2}\Delta_0 & 0 & 0 & -i\lambda-k_y\\
    0 & \frac{1}{2}\Delta_0 & -i\lambda-k_y & 0
    \end{pmatrix},
\end{align}
and the low-energy effective Hamiltonian is
\begin{align}
    H_{\textnormal{eff}}
    &=
    \begin{pmatrix}
    \langle\Psi_\alpha^o|\Delta H|\Psi_\alpha^o\rangle & \langle\Psi_\alpha^o|\Delta H|\Psi_\beta^o\rangle & \langle\Psi_\alpha^o|\Delta H|\Psi_\gamma^o\rangle & \langle\Psi_\alpha^o|\Delta H|\Psi_\delta^o\rangle\\
    \langle\Psi_\beta^o|\Delta H|\Psi_\alpha^o\rangle & \langle\Psi_\beta^o|\Delta H|\Psi_\beta^o\rangle & \langle\Psi_\beta^o|\Delta H|\Psi_\gamma^o\rangle & \langle\Psi_\beta^o|\Delta H|\Psi_\delta^o\rangle\\
    \langle\Psi_\gamma^o|\Delta H|\Psi_\alpha^o\rangle & \langle\Psi_\gamma^o|\Delta H|\Psi_\beta^o\rangle & \langle\Psi_\gamma^o|\Delta H|\Psi_\gamma^o\rangle & \langle\Psi_\gamma^o|\Delta H|\Psi_\delta^o\rangle\\
    \langle\Psi_\delta^o|\Delta H|\Psi_\alpha^o\rangle & \langle\Psi_\delta^o|\Delta H|\Psi_\beta^o\rangle & \langle\Psi_\delta^o|\Delta H|\Psi_\gamma^o\rangle & \langle\Psi_\delta^o|\Delta H|\Psi_\delta^o\rangle
    \end{pmatrix}.
\end{align}
The overlap integral in the eigenbasis of $|\nu_i\rangle$'s is:
\begin{align}
    \langle\nu_3|\Delta H_1^\dagger|\nu_3\rangle &= i\lambda-k_y & \langle\nu_3|\Delta H_1^\dagger|\nu_4\rangle &= -\frac{1}{2}\Delta_0 & \langle\nu_4|\Delta H_1^\dagger|\nu_4\rangle &= i\lambda+k_y\\
    \langle\nu_3|\Delta H_0|\nu_3\rangle &=-k_y &
    \langle\nu_3|\Delta H_0|\nu_4\rangle &=0 &
    \langle\nu_4|\Delta H_0|\nu_4\rangle &=k_y \\
    \langle\nu_3|\Delta H_1|\nu_3\rangle &= -i\lambda-k_y & \langle\nu_3|\Delta H_1|\nu_4\rangle &= \frac{1}{2}\Delta_0 & \langle\nu_4|\Delta H_1|\nu_4\rangle &= -i\lambda+k_y.
\end{align}
And it is useful to note the following sums (and that $\gamma_j=\alpha_j$ and $\delta_j=\beta_j$) :
\begin{align}
    &\sum_j \alpha_j \alpha_{j} = \sum_j \beta_j \beta_{j} = 1&
    &\sum_j \alpha_j \beta_{j} = \frac{1}{\sqrt{5}}\\
    &\sum_j \alpha_j \alpha_{j+1} = -\frac{1}{2} &
    &\sum_j \beta_j \beta_{j+1} = -\frac{4}{5} \\
    &\sum_j \alpha_j \beta_{j+1} = -\frac{2}{\sqrt{5}} &
    &\sum_j \beta_j \alpha_{j+1} = -\frac{1}{2\sqrt{5}}.
\end{align}
The matrix elements are thus calculated as:
\begin{align}
    \langle\Psi_\alpha^o|\Delta H|\Psi_\alpha^o\rangle
    &= \sum_j \alpha_j \langle\nu_3|(\alpha_{j-1}\Delta H_1^\dagger+\alpha_j\Delta H_0+\alpha_{j+1}\Delta H_1)|\nu_3\rangle \nonumber \\
    &=\sum_j\alpha_j\alpha_{j+1}(i\lambda-k_y)+\alpha_j\alpha_{j}(-k_y)+\alpha_j\alpha_{j+1}(-i\lambda-k_y) \nonumber\\
    &=(-k_y)\sum_j \alpha_j\alpha_{j} + 2\alpha_j\alpha_{j+1} \nonumber\\
    &=(-k_y)(1-2\times\frac{1}{2}) \nonumber\\
    &=0
\end{align}
\begin{align}
    \langle\Psi_\alpha^o|\Delta H|\Psi_\beta^o\rangle
    &= \langle\Psi_\alpha|\Delta H(\frac{\sqrt{5}}{2}|\Psi_\beta\rangle - \frac{1}{2}|\Psi_\alpha\rangle) \nonumber\\
    &=\frac{\sqrt{5}}{2}\sum_j \alpha_j \langle\nu_3|(\beta_{j-1}\Delta H_1^\dagger+\beta_j\Delta H_0+\beta_{j+1}\Delta H_1)|\nu_3
    \rangle - \frac{1}{2}\times 0 \nonumber\\
    &=\frac{\sqrt{5}}{2}\sum_j\alpha_{j+1}\beta_{j}(i\lambda-k_y)+\alpha_j\beta_{j}(-k_y)+\alpha_j\beta_{j+1}(-i\lambda-k_y) \nonumber\\
    &=\frac{\sqrt{5}}{2} \left[ (-\frac{1}{2\sqrt{5}})(i\lambda-k_y)+\frac{1}{\sqrt{5}}(-k_y)+(-\frac{2}{\sqrt{5}})(-i\lambda-k_y)  \right] \nonumber\\
    &=-\frac{1}{4}i\lambda + \frac{1}{4}k_y -\frac{1}{2}k_y + i\lambda +k_y \nonumber\\
    &= \frac{3}{4}k_y + \frac{3}{4}i\lambda
\end{align}
and
\begin{align}
    \langle\Psi_\alpha^o|\Delta H|\Psi_\gamma^o\rangle
    &= \sum_j \alpha_j \langle\nu_3|(\alpha_{j-1}\Delta H_1^\dagger+\alpha_j\Delta H_0+\alpha_{j+1}\Delta H_1)|\nu_4\rangle \nonumber\\
    &=\sum_j\alpha_{j+1}\alpha_{j}(-\frac{1}{2}\Delta_0)+\alpha_j\alpha_{j}(0)+\alpha_j\alpha_{j+1}(\frac{1}{2}\Delta_0) \nonumber\\
    &=0
\end{align}
\begin{align}
    \langle\Psi_\alpha^o|\Delta H|\Psi_\delta^o\rangle
    &= \langle\Psi_\alpha|\Delta H (\frac{\sqrt{5}}{2}|\Psi_\delta\rangle - \frac{1}{2}|\Psi_\gamma\rangle) \nonumber\\
    &=\frac{\sqrt{5}}{2}\sum_j \alpha_j \langle\nu_3|(\beta_{j-1}\Delta H_1^\dagger+\beta_j\Delta H_0+\beta_{j+1}\Delta H_1)|\nu_4\rangle- \frac{1}{2}\times 0 \nonumber\\
    &=\frac{\sqrt{5}}{2}\sum_j\alpha_{j+1}\beta_{j}(-\frac{1}{2}\Delta_0)+\alpha_j\beta_{j}(0)+\alpha_j\beta_{j+1}(\frac{1}{2}\Delta_0) \nonumber\\
    &=\frac{\sqrt{5}}{2} \left[ (-\frac{1}{2\sqrt{5}})(-\frac{1}{2}\Delta_0) +  (-\frac{2}{\sqrt{5}})(\frac{1}{2}\Delta_0) \right] \nonumber\\
    &=-\frac{3}{8}\Delta_0
\end{align}
Moving on to the next line in $H_{\textnormal{eff}}$:
\begin{align}
    \langle\Psi_\beta^o|\Delta H|\Psi_\beta^o\rangle
    &= (\frac{\sqrt{5}}{2}\langle\Psi_\beta| - \frac{1}{2}\langle\Psi_\alpha|) \Delta H(\frac{\sqrt{5}}{2}|\Psi_\beta\rangle - \frac{1}{2}|\Psi_\alpha\rangle) \nonumber\\
    &=\frac{5}{4}\langle\Psi_\beta|\Delta H|\Psi_\beta\rangle -\frac{\sqrt{5}}{4}\langle\Psi_\alpha|\Delta H|\Psi_\beta\rangle-\frac{\sqrt{5}}{4}\langle\Psi_\beta|\Delta H|\Psi_\alpha\rangle +\frac{1}{4}\langle\Psi_\alpha|\Delta H|\Psi_\alpha\rangle \nonumber\\
    &=\frac{5}{4}\sum_j \beta_j \langle\nu_3|(\beta_{j-1}\Delta H_1^\dagger+\beta_j\Delta H_0+\beta_{j+1}\Delta H_1)|\nu_3\rangle- \frac{1}{2}(\frac{3}{4}k_y + \frac{3}{4}i\lambda)- \frac{1}{2}(\frac{3}{4}k_y - \frac{3}{4}i\lambda) +0 \nonumber\\
    &=-\frac{3}{4} k_y+\frac{5}{4}\sum_j\beta_j\beta_{j+1}(i\lambda-k_y)+\beta_j\beta_{j}(-k_y)+\beta_j\beta_{j+1}(-i\lambda-k_y) \nonumber\\
    &=-\frac{3}{4} k_y+\frac{5}{4} \left[ (-\frac{4}{5})(i\lambda-k_y)+(-k_y)+(-\frac{4}{5})(-i\lambda-k_y) \right] \nonumber\\
    &=0
\end{align}
\begin{align}
    \langle\Psi_\beta^o|\Delta H|\Psi_\gamma^o\rangle
    &= (\frac{\sqrt{5}}{2}\langle\Psi_\beta| - \frac{1}{2}\langle\Psi_\alpha|) \Delta H|\Psi_\gamma\rangle \nonumber\\
    &=\frac{\sqrt{5}}{2}\langle\Psi_\beta|\Delta H|\Psi_\gamma\rangle -\frac{1}{2}\langle\Psi_\alpha|\Delta H|\Psi_\gamma\rangle \nonumber\\
    &=\frac{\sqrt{5}}{2}\sum_j \beta_j \langle\nu_3|(\alpha_{j-1}\Delta H_1^\dagger+\alpha_j\Delta H_0+\alpha_{j+1}\Delta H_1)|\nu_4\rangle - 0 \nonumber\\
    &=\frac{\sqrt{5}}{2}\sum_j\beta_{j+1}\alpha_{j}(-\frac{1}{2}\Delta_0)+\beta_j\alpha_{j}(0)+\beta_{j}\alpha_{j+1}(\frac{1}{2}\Delta_0) \nonumber\\
    &=\frac{\sqrt{5}}{2} \left[ (-\frac{2}{\sqrt{5}})(-\frac{1}{2}\Delta_0) + (-\frac{1}{2\sqrt{5}}) (\frac{1}{2}\Delta_0) \right] \nonumber\\
    &=\frac{3}{8}\Delta_0
\end{align}
\begin{align}
    \langle\Psi_\beta^o|\Delta H|\Psi_\delta^o\rangle
    &= (\frac{\sqrt{5}}{2}\langle\Psi_\beta| - \frac{1}{2}\langle\Psi_\alpha|) \Delta H(\frac{\sqrt{5}}{2}|\Psi_\delta\rangle - \frac{1}{2}|\Psi_\gamma\rangle) \nonumber\\
    &=\frac{5}{4}\langle\Psi_\beta|\Delta H|\Psi_\delta\rangle -\frac{\sqrt{5}}{4}\langle\Psi_\beta|\Delta H|\Psi_\gamma\rangle-\frac{\sqrt{5}}{4}\langle\Psi_\alpha|\Delta H|\Psi_\delta\rangle +\frac{1}{4}\langle\Psi_\alpha|\Delta H|\Psi_\gamma\rangle \nonumber\\
    &=\frac{5}{4}\sum_j \beta_j \langle\nu_3|(\beta_{j-1}\Delta H_1^\dagger+\beta_j\Delta H_0+\beta_{j+1}\Delta H_1)|\nu_4\rangle- \frac{1}{2}(\frac{3}{8}\Delta_0)- \frac{1}{2}(-\frac{3}{8}\Delta_0) +0 \nonumber\\
    &=\frac{5}{4}\sum_j\beta_{j+1}\beta_{j}(-\frac{1}{2}\Delta_0)+\beta_j\beta_{j}(0)+\beta_j\beta_{j+1}(\frac{1}{2}\Delta_0) \nonumber\\
    &=0
\end{align}
Moving on to the next line of $H_{\textnormal{eff}}$:
\begin{align}
    \langle\Psi_\gamma^o|\Delta H|\Psi_\gamma^o\rangle
    &= \sum_j \alpha_j \langle\nu_4|(\alpha_{j-1}\Delta H_1^\dagger+\alpha_j\Delta H_0+\alpha_{j+1}\Delta H_1)|\nu_4\rangle \nonumber\\
    &=\sum_j\alpha_{j+1}\alpha_{j}(i\lambda+k_y)+\alpha_j\alpha_{j}(k_y)+\alpha_j\alpha_{j+1}(-i\lambda+k_y) \nonumber\\
    &=k_y\sum_j \alpha_j\alpha_{j} + 2\alpha_j\alpha_{j+1} \nonumber\\
    &=k_y(1-2\times\frac{1}{2}) \nonumber\\
    &=0
\end{align}
\begin{align}
    \langle\Psi_\gamma^o|\Delta H|\Psi_\delta^o\rangle
    &= \langle\Psi_\gamma|\Delta H(\frac{\sqrt{5}}{2}|\Psi_\delta\rangle - \frac{1}{2}|\Psi_\gamma\rangle) \nonumber\\
    &=\frac{\sqrt{5}}{2}\sum_j \alpha_j \langle\nu_4|(\beta_{j-1}\Delta H_1^\dagger+\beta_j\Delta H_0+\beta_{j+1}\Delta H_1)|\nu_4
    \rangle - \frac{1}{2}\times 0 \nonumber\\
    &=\frac{\sqrt{5}}{2}\sum_j\alpha_{j+1}\beta_{j}(i\lambda+k_y)+\alpha_j\beta_{j}(k_y)+\alpha_j\beta_{j+1}(-i\lambda+k_y) \nonumber\\
    &=\frac{\sqrt{5}}{2} \left[ (-\frac{1}{2\sqrt{5}})(i\lambda+k_y)+\frac{1}{\sqrt{5}}(k_y)+(-\frac{2}{\sqrt{5}})(-i\lambda+k_y)  \right] \nonumber\\
    &=-\frac{1}{4}i\lambda - \frac{1}{4}k_y +\frac{1}{2}k_y + i\lambda -k_y \nonumber\\
    &= -\frac{3}{4}k_y + \frac{3}{4}i\lambda
\end{align}
\begin{align}
    \langle\Psi_\delta^o|\Delta H|\Psi_\delta^o\rangle
    &= (\frac{\sqrt{5}}{2}\langle\Psi_\delta| - \frac{1}{2}\langle\Psi_\gamma|) \Delta H(\frac{\sqrt{5}}{2}|\Psi_\delta\rangle - \frac{1}{2}|\Psi_\gamma\rangle) \nonumber\\
    &=\frac{5}{4}\langle\Psi_\delta|\Delta H|\Psi_\delta\rangle -\frac{\sqrt{5}}{4}\langle\Psi_\delta|\Delta H|\Psi_\gamma\rangle-\frac{\sqrt{5}}{4}\langle\Psi_\gamma|\Delta H|\Psi_\delta\rangle +\frac{1}{4}\langle\Psi_\gamma|\Delta H|\Psi_\gamma\rangle \nonumber\\
    &=\frac{5}{4}\sum_j \beta_j \langle\nu_4|(\beta_{j-1}\Delta H_1^\dagger+\beta_j\Delta H_0+\beta_{j+1}\Delta H_1)|\nu_4\rangle- \frac{1}{2}(-\frac{3}{4}k_y - \frac{3}{4}i\lambda)- \frac{1}{2}(-\frac{3}{4}k_y + \frac{3}{4}i\lambda) +0 \nonumber\\
    &=\frac{3}{4} k_y+\frac{5}{4}\sum_j\beta_j\beta_{j+1}(i\lambda+k_y)+\beta_j\beta_{j}(k_y)+\beta_j\beta_{j+1}(-i\lambda+k_y) \nonumber\\
    &=\frac{3}{4} k_y+\frac{5}{4} \left[ (-\frac{4}{5})(i\lambda+k_y)+(k_y)+(-\frac{4}{5})(-i\lambda+k_y) \right] \nonumber\\
    &=0
\end{align}
Collecting all terms together, the effective Hamiltonian is
\begin{align}
   H_{\textnormal{eff}} &=
   \begin{pmatrix}
    0 & \frac{3}{4}k_y+\frac{3}{4}i\lambda & -\frac{3}{8}\Delta_0 & 0\\
    \frac{3}{4}k_y-\frac{3}{4}i\lambda & 0 & 0 & \frac{3}{8}\Delta_0\\
    -\frac{3}{8}\Delta_0 & 0 & 0 & -\frac{3}{4}k_y + \frac{3}{4}i\lambda\\
    0 & \frac{3}{8}\Delta_0 & -\frac{3}{4}k_y - \frac{3}{4}i\lambda & 0
    \end{pmatrix}
\end{align}
By taking $H_{\textnormal{eff}} (k_y\rightarrow -i\partial_y, \lambda \rightarrow y/R)$, $H_{\textnormal{eff}}$ becomes
\begin{align}
   H_{\textnormal{eff}} &= \frac{3}{4}
   \begin{pmatrix}
    0 & -i\partial_y+i\frac{1}{R}y & -\frac{1}{2}\Delta_0 & 0\\
    -i\partial_y-i\frac{1}{R}y & 0 & 0 & \frac{1}{2}\Delta_0\\
    -\frac{1}{2}\Delta_0 & 0 & 0 & i\partial_y+i\frac{1}{R}y\\
    0 & \frac{1}{2}\Delta_0 & i\partial_y-i\frac{1}{R}y & 0
    \end{pmatrix}
\end{align}
and squaring both sides of the time-independent Schrödinger equation yields
\begin{equation}
\begin{aligned}
    a_{11} &= (0 , -i\partial_y+i\frac{1}{R}y, -\frac{1}{2}\Delta_0 , 0) \cdot (0 , -i\partial_y-i\frac{1}{R}y, -\frac{1}{2}\Delta_0 , 0)\\
    &= (-\partial_y+\frac{1}{R}y)(\partial_y+\frac{1}{R}y)+\frac{1}{4}\Delta_0^2\\
    &=-\partial_y^2-\frac{1}{R}\left( y\partial_y - \partial_yy \right)+\frac{1}{R^2}y^2 +\frac{1}{4}\Delta_0^2\\
    &=-\partial_y^2-\frac{1}{R}+\frac{1}{R^2}y^2+\frac{1}{4}\Delta_0^2\\
    a_{14} &= (0 , -i\partial_y+i\frac{1}{R}y, -\frac{1}{2}\Delta_0 , 0) \cdot (0 , \frac{1}{2}\Delta_0, i\partial_y+i\frac{1}{R}y , 0)\\
    &= \frac{1}{2}\Delta_0 (-i\partial_y+i\frac{1}{R}y-i\partial_y-i\frac{1}{R}y)\\
    &=-i\Delta_0\partial_y
\end{aligned}
\end{equation}

And the remaining elements are calculated in a similar way. The resultant equation is
\begin{align}\frac{9}{16}&
    \begin{pmatrix}
    \epsilon_{y,-} & 0 & 0 & -i\partial_y \Delta_0\\
    0 & \epsilon_{y,+} & i\partial_y \Delta_0 & 0\\
    0 & i\partial_y \Delta_0 & \epsilon_{y,+} & 0\\
    -i\partial_y \Delta_0 & 0 & 0 & \epsilon_{y,-}
    \end{pmatrix} \psi = E^2 \psi,
\end{align}
where $\epsilon_{y,\pm} = -\partial_y^2+\frac{1}{R^2}y^2\pm \frac{1}{R}+\frac{1}{4}\Delta_0^2$.

The eigenstates of this low-energy effective Hamiltonian may then be expressed as $|\mu_i \rangle = \chi_i^{\top} |\psi_i \rangle$, where $\chi_i$ satisfies the constraints of the Hamiltonian matrix representation, and $|\psi_i\rangle = |\psi_i(y)\rangle$ is a spatially-varying scalar function satisfying each of the differential equations contained in the Hamiltonian. We first identify $\{\chi_i \}$ ($i \in \{1,2,3,4\}$) as eigenvectors of $\tau_2\sigma_2$, yielding $|\mu_i\rangle$ which take the forms
\begin{equation}
\begin{aligned}
\label{eigenstates}
    |\mu_1\rangle &= \frac{1}{\sqrt{2}}(1,0,0,1)^\top |\psi_1\rangle&
    |\mu_2\rangle &= \frac{1}{\sqrt{2}}(0,-1,1,0)^\top |\psi_2\rangle\\
    |\mu_3\rangle &= \frac{1}{\sqrt{2}}(-1,0,0,1)^\top |\psi_3\rangle&
    |\mu_4\rangle &= \frac{1}{\sqrt{2}}(0,1,1,0)^\top |\psi_4\rangle
\end{aligned}
\end{equation}

\begin{figure}[t]
\centering
\includegraphics[width=1\textwidth]{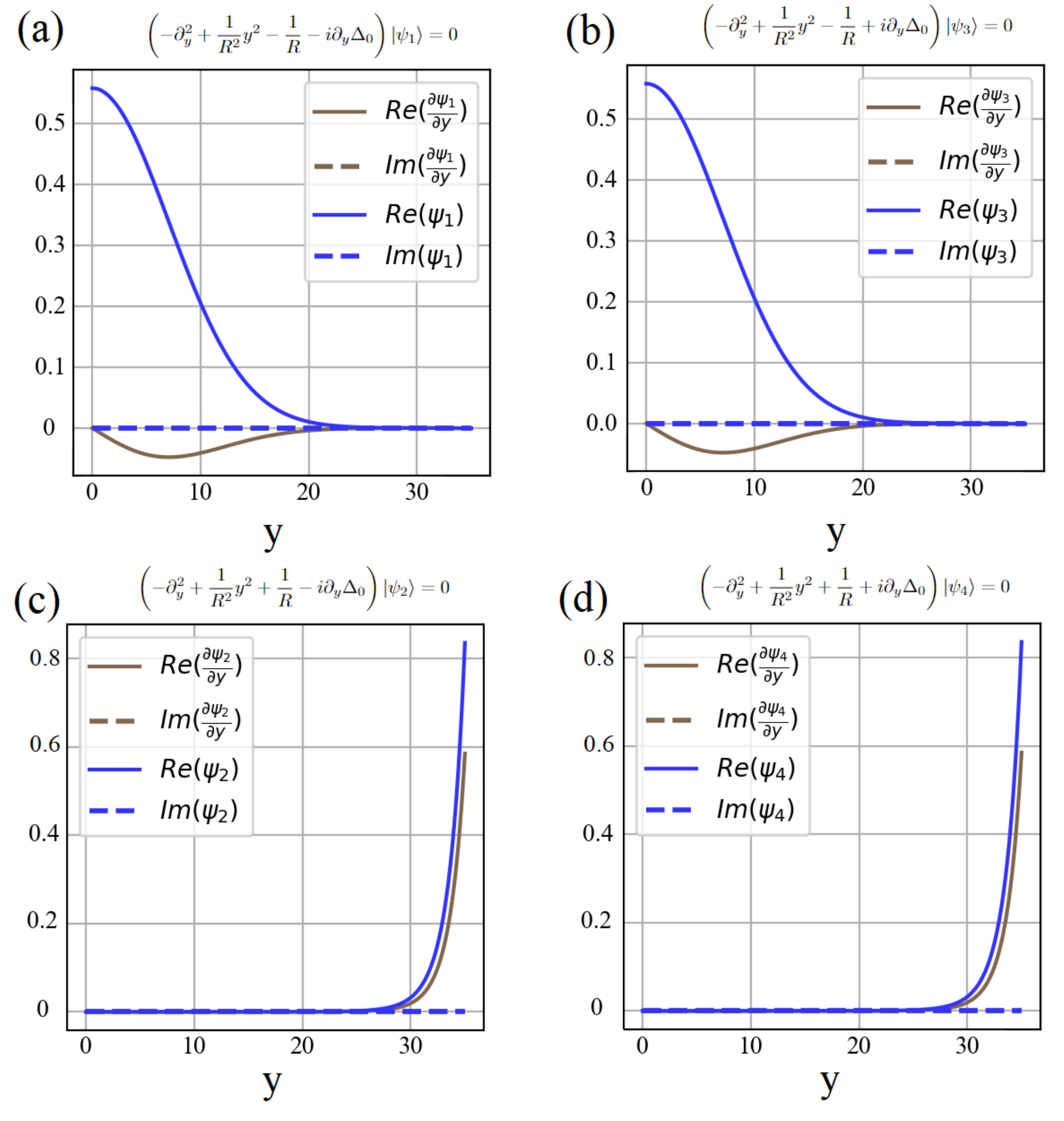}
\caption{Numerical solutions for Eq.~\eqref{DE1}-Eq.\eqref{DE4} \bl{with $R=50$ and $\Delta_0=0.1$}. In (a), the real and imaginary components of \bl{normalized} $\psi_1$ and $\frac{\partial \psi_1}{\partial y}$ are plotted. Similarly, (b), (c) and (d) show the real and imaginary components $\psi_3$, $\psi_2$, $\psi_4$, and their correponding partial derivatives, respectively. (a) and (b) show that Eq.~\eqref{DE1} and Eq.~\eqref{DE3}, \bl{which are the $-\frac{1}{R}$ equations with $\pm i\partial_y \Delta_0$ term}, have localized solutions at $y=0$, and (c)(d), \bl{which are the $+\frac{1}{R}$ equations with $\pm i\partial_y \Delta_0$ term}, show that Eq.~\eqref{DE2} and Eq.~\eqref{DE4} have no localized solutions as the solution grows exponentially with increasing $y$. }
\label{fig:localization}
\end{figure}

These states serve as zero-energy eigenstates of the effective low-energy Hamiltonian if the following differential equations are satisfied by the $\{\psi_i(y) \}$ (omitting $\frac{1}{4}\Delta_0^2$ as higher order terms):
\begin{align}
    \label{DE1}
    \left(-\partial_y^2+\frac{1}{R^2}y^2-\frac{1}{R}-i\partial_y \Delta_0\right)|\psi_1\rangle &= 0\\
    \label{DE2}
    \left(-\partial_y^2+\frac{1}{R^2}y^2+\frac{1}{R}-i\partial_y \Delta_0\right)|\psi_2\rangle &= 0\\
    \label{DE3}
    \left(-\partial_y^2+\frac{1}{R^2}y^2-\frac{1}{R}+i\partial_y \Delta_0\right)|\psi_3\rangle &= 0\\
    \label{DE4}
    \left(-\partial_y^2+\frac{1}{R^2}y^2+\frac{1}{R}+i\partial_y \Delta_0\right)|\psi_4\rangle &= 0.
\end{align}
Note that by setting $\Delta_0=0$ we recover the differential equation in the two-band Hopf insulator which only has solution for $-\frac{1}{R}$, so that we can take this solution and treat $\pm i\partial _y \Delta_0$ as a perturbation term so that the resultant wavefunction is still strongly localized at $y=0$, as shown in Fig.~\ref{fig:localization}(a) and (b). On the other hand, for the $+\frac{1}{R}$ part which has no normalizable solution in the two-band Hopf insulator case, there are still no normalizable solutions with the addition of the $\pm i\partial _y \Delta_0$ term. The solution instead grows exponentially with increasing $y$ as shown in Fig.~\ref{fig:localization}(c) and (d).

Therefore, there are zero-energy eigenstates at the inner boundary of the form,
\begin{align}
    |0\rangle_R   &=|\psi_3(y)\rangle
                \begin{pmatrix}
                1  \\
                0 \\
                0\\
                -1\\
                \end{pmatrix}\\
    |0\rangle_I   &=|\psi_1(y)\rangle
                \begin{pmatrix}
                i  \\
                0 \\
                0\\
                i\\
                \end{pmatrix}.
\end{align}

Note that the differential equations~\eqref{DE1}-\eqref{DE4} are obtained from $|\mu_{1,2,3,4}\rangle$, and would be identical if $i|\mu_{1,2,3,4}\rangle$ were used instead. Therefore, by including a factor of $i$ in $|\mu_1\rangle$, we obtain $|0\rangle_{R(I)}$ as the purely real (imaginary), $\mathcal{C}'$-invariant Majorana zero-modes. More generally, from numerical diagonalization, we expect two states $|0\rangle_{\pm} = |0\rangle_R \pm |0\rangle_I$. Correspondingly, $|0\rangle_{R(I)} = |0\rangle_{+} + (-)|0\rangle_{-}$.

The above analysis is carried out for $x>0$ i.e. the inner boundary. As for $x<0$ (the outer boundary), the essential difference is that we swap $H_1$ with $H_1^\dagger$ and $H_2$ with $H_2^\dagger$ in Eqn~\eqref{eqn:main}:
\begin{equation}
    \begin{pmatrix}
    H_0 & H_1^\dagger & H_2^\dagger & 0 & 0 & 0 & \cdots \\
    H_1 & H_0 & H_1^\dagger & H_2^\dagger & 0 & 0 & \cdots \\
    H_2 & H_1 & H_0 & H_1^\dagger & H_2^\dagger & 0 & \cdots \\
    0 & H_2 & H_1 & H_0 & H_1^\dagger & H_2^\dagger & \cdots \\
    \vdots & \vdots & \vdots & \vdots & \vdots & \vdots & \ddots \\
    \end{pmatrix}
    \begin{pmatrix}
    \psi_{-1} \\
    \psi_{-2} \\
    \psi_{-3} \\
    \psi_{-4} \\
    \vdots
    \end{pmatrix}
    =
    E
    \begin{pmatrix}
    \psi_{-1} \\
    \psi_{-2} \\
    \psi_{-3} \\
    \psi_{-4} \\
    \vdots
    \end{pmatrix}
\end{equation}
and repeat the same calculation. This time we get
\begin{align}\frac{9}{16}&
    \begin{pmatrix}
    \epsilon_{y,-} & 0 & 0 & i\Delta_0\partial_y \\
    0 & \epsilon_{y,+} & -i\Delta_0\partial_y & 0\\
    0 & -i\Delta_0\partial_y & \epsilon_{y,+} & 0\\
    i\Delta_0\partial_y & 0 & 0 & \epsilon_{y,-}
    \end{pmatrix} \psi = E^2 \psi,
\end{align}
where $\epsilon_{y,\pm} = -\partial_y^2+\frac{1}{R^2}y^2\pm \frac{1}{R}$.

Subsequently we will reach the same differential equations \eqref{DE1} and \eqref{DE3}, again with eigenstates $|\mu_1\rangle$ and $|\mu_3\rangle$ in Eqn~\eqref{eigenstates}. Hence, we similarly find two additional zero-energy states of the form,

\begin{align}
    |\bar{0}\rangle_R   &=|\psi_1(y)\rangle
                \begin{pmatrix}
                1  \\
                0 \\
                0\\
                -1\\
                \end{pmatrix}\\
    |\bar{0}\rangle_I   &=|\psi_3(y)\rangle
                \begin{pmatrix}
                i  \\
                0 \\
                0 \\
                i \\
                \end{pmatrix}.
\end{align}

Thus, if the system has $\mathcal{C}'$ symmetry alone, there will be no localized Majorana zero modes which are equal-weight superposition of particle and hole components.

In the case of the skyrmion phase in the main text, $|0\rangle_{R(I)}$ is further constrained by $|\bar{0}\rangle_{I(R)}$, as the linear combinations $|0\rangle_{R(I)} \pm |\bar{0}\rangle_{I(R)}$ must also respect this particle-hole symmetry. These constraints are severe enough to enforce a cross-like structure on each of the four localized eigenstates of the Hamiltonian at approximately zero energy, such that each consists of two Majorana zero-modes \textit{which are now isolated from one another} by a $\pi/2$ phase in the complex plane, rather than being isolated from one another by separation over real-space.

\end{document}